\documentclass[12pt]{article}
\pdfoutput=1
\usepackage{rsfso}
\usepackage[utf8]{inputenc}
\usepackage[left=2.55cm, right=2.55cm, top=2.55cm, bottom=2.55cm]{geometry}
\usepackage{tikz}
\usepackage{tikz-feynman}
\usepackage{contour}
\usepackage{amsmath,amssymb,amsbsy}
\usepackage{slashed}
\usepackage{xcolor}
\usepackage{graphicx}
\usepackage{tikz}
\usepackage{amsmath}
\usepackage{cases}
\usepackage{tikz-feynman}
\usepackage{contour}
\usepackage{url}
\usepackage{cancel}
\usepackage{cite}
\usepackage[colorlinks=true,allcolors=blue,pdfborder={0 0 0},linktocpage=false,pdfencoding=auto]{hyperref}
\usepackage{tabularx,booktabs}
\usepackage{multicol}
\usepackage{feynmp}
\usepackage{units}
\usepackage{xspace}
\usepackage[labelfont=bf]{caption}
\usepackage[section]{placeins}
\usepackage{subcaption}
\usepackage{soul}
\usepackage{bbold}
\usepackage{parskip}
\usepackage{float}
\usepackage{tabulary}

\definecolor{darkred}{rgb}{0.6,0,0}
\definecolor{darkpurple}{rgb}{0.5,0,0.5}

\newcommand{\beqn}{\begin{eqnarray}}
\newcommand{\eeqn}{\end{eqnarray}}
\def\non{\nonumber\\}

\def\gws{{\rm gravitational waves}~}
\def\gw{{\rm gravitational wave}~}
\def\h{\rm{hid}}

\begin{document}
\author{Jinzheng Li$^a$\footnote{\href{mailto:li.jinzh@northeastern.edu}{li.jinzh@northeastern.edu}} ~and
Pran Nath$^a$\footnote{\href{mailto:p.nath@northeastern.edu}{p.nath@northeastern.edu}}\\
$^{a}$\textit{\normalsize Department of Physics, Northeastern University, Boston, MA 02115-5000, USA} }

\title{\vspace{-2cm}\begin{flushright}
\end{flushright}
\vspace{1cm}
 \bf\large

Supercooled Phase Transitions: Why Thermal History of Hidden Sector Matters in Analysis of Pulsar Timing Array Signals}
 \vspace{0.0cm}

\date{}
\maketitle
\begin{abstract}

The detection of a gravitational wave background in the nano-Hertz frequency range
from Pulsar Timing Array (PTA)  observations offers new insights into evolution of the early universe. In this work we analyze  \gw data from
 PPTA, EPTA, and NANOGrav, as arising from a supercooled first-order phase transition within a hidden sector, characterized by a broken $U(1)_X$ gauge symmetry. {Several previous works have discussed challenges} in producing 
observable {PTA signal} from supercooled phases transitions. We discuss these
challenges and show how they are overcome by inclusion in part of the proper thermal history of the hidden and the visible sectors.
 The analysis of this work demonstrates that thermal histories of hidden and visible sectors  profoundly influence the gravitational wave power spectrum, an aspect not previously explored in the literature.  Further, the analysis of this work suggests that supercooled phase transitions  not only align with the Pulsar Timing Array observations but also show promise for gravitational wave detection by future gravitational wave detectors. 
 Our analysis shows that the dominant contribution to the \gw power spectrum for PTA signal comes from bubble collision while the sound wave and turbulence contributions are highly  suppressed. It is also found that all the PTA events are of detonation type while deflagration and hybrid events are absent. 
The analysis presented in this work provides a robust framework for further investigations on the origin of gravitational wave power spectrum in the early universe and for their experimental observation in the future.
\end{abstract}

\numberwithin{equation}{section}
\newpage
{  \hrule height 0.4mm \hypersetup{colorlinks=black,linktocpage=true} 
\tableofcontents
\vspace{0.5cm}
 \hrule height 0.4mm} 
\section{Introduction}
\label{sec:intro}
Recent observations by various Pulsar Timing Array (PTA) collaborations, including 
PPTA~\cite{Reardon:2023gzh},  EPTA~\cite{EPTA:2023fyk}
and NANOGrav~\cite{NANOGrav:2023gor}, have detected a gravitational wave (GW) background signal in the nano-Hertz frequency range. These observations have sparked considerable interest
{as they provide}
  insights into the early {history of the} universe. Among the possible explanations, supercooled first-order phase transitions (FOPTs) have emerged as a particularly attractive scenario~\cite{Athron:2023mer,Kawana:2022fum,Lewicki:2022pdb,Hawking:1981fz,Ghosh:2023aum,Athron:2023rfq,Madge:2023dxc,Fujikura:2023lkn,Athron:2023xlk,Sagunski:2023ynd,Nakai:2020oit,Ellis:2020nnr,Kobakhidze:2017mru,Gouttenoire:2023naa,Freese:2022qrl,Fairbairn:2019xog,Wang:2020jrd,Megevand:2016lpr,Leitao:2015fmj,Ellis:2019oqb}. Here a  sufficiently strong FOPT that remains trapped in a metastable false vacuum until very low temperatures could in principle generate   \gws with amplitudes and spectral shapes compatible with the current PTA observations.
However, recent studies attempting to fit PTA signals using power-law indices and normalizations 
~\cite{Freese:2022qrl,Bringmann:2023opz,NANOGrav:2020bcs,Goncharov:2021oub,EPTA:2021crs,Blasi:2020mfx,Benetti:2021uea,Vaskonen:2020lbd,Buchmuller:2020lbh} 
consistently find that the nucleation parameter $\beta/H$ must be less than approximately $10^2$. This small value of $\beta/H$ indicates a strongly supercooled phase transition. In contrast, moderately supercooled transitions typically yield $\beta/H > 100$~\cite{Bringmann:2023iuz,Kawana:2022fum,Wang:2020jrd}. The PTA observations thus suggest that if the detected \gws originate from a cosmic FOPT, they likely arise from a supercooled scenario. However, as noted in~\cite{Athron:2023mer}, several challenges emerge when studying strongly supercooled phase transitions, potentially invalidating some previously proposed models. Our careful analysis reveals that a supercooled FOPT in a hidden sector remains a viable explanation for the PTA signal, even after accounting for these challenges.

In this work, we explore a supercooled first-order phase transition (FOPT) in a hidden sector with a spontaneously broken $U(1)_X$ gauge symmetry, comprising a hidden gauge boson, a dark scalar field, and a dark fermion. Similar studies on hidden sector phase transitions have been conducted in Refs.~\cite{Breitbach:2018ddu,Fairbairn:2019xog,Ertas:2021xeh,Freese:2022qrl,Bringmann:2023iuz,Banik:2024zwj,Jaeckel:2016jlh, Schwaller:2015tja,Li:2023bxy,Ghosh:2023aum,DiBari:2021dri,Chen:2023rrl,Gouttenoire:2023naa,Kanemura:2023jiw}. However, these studies generally use the inverse timescale of the transition $\beta$, which has been shown to yield significant errors in supercooled phase transition analyses; instead, the mean bubble separation $R$ should be used \cite{Athron:2023xlk,Athron:2023mer}. We carefully compare the results obtained using these two approaches and demonstrate that using $R_*$ not only provides more accurate results but also validates previously excluded parameters that could explain the PTA signal.

A central piece of our analysis focuses on the temperature ratio $\xi$ between the hidden and visible sector temperatures, which plays a key role in the hidden sector phase transitions. This ratio influences two parameters: the effective number of relativistic species,  {$N_{\rm eff}$}, and the transition strength, $\alpha_{\rm tot}$. When explaining PTA signals while satisfying Big Bang Nucleosynthesis (BBN) cosmological constraints \cite{Cyburt:2015mya}, one finds a strong tension between these parameters.
Here we explore a decaying hidden sector as a possible solution, as noticed earlier 
in \cite{Bringmann:2023opz}.
Through solutions of coupled Boltzmann equations, we demonstrate how hidden sector transitions from relativistic to non-relativistic before BBN, simultaneously satisfying both $\Delta N_{\rm eff}$ and relic density constraints while generating a sufficiently large $\alpha_{\rm tot}$ to explain PTA signal.

The evolution of the temperature ratio proves to be critical in analyzing the supercooled phase transition. {Thus the  computation of the percolation temperature $T_{h,p}$ and the mean bubble separation $R_*$, involves evaluation} of integrals from the critical temperature $T_{h,c}$ to  the percolation temperature $T_{h,p}$. In a supercooled phase transition, we typically observe $T_{h,p}/T_{h,c} \ll 1$, indicating a large temperature interval. Consequently, the evolution of the Hubble parameter $H(T_h)$ during this period becomes significant, as the temperature ratio evolution {$\xi(T_v)= T_v/T_h$  or $\zeta(T_h)= T_h/T_v$} 
strongly influences it. Our analysis shows 
that {the gravitational power spectrum is sensitively dependent on $\xi(t_p)(\zeta(t_p))$ and   thus different thermal evolutions can alter \gw power spectrum by up to four orders of magnitude}.
Finally, we exhibit benchmark models that generate gravitational waves with sufficient strength to explain the PTA signal while satisfying all other cosmological constraints. Further, some of these models predict signals within the detection range of future space-based gravitational wave detectors. These  detectors could observe the supercooled FOPT 
signals above the NanoGrav, EPTA, PPTA  frequency range as further test of the models
discussed here.

The rest of this paper is organized as follows. In section~\ref{sec:model}, we discuss the hidden $U(1)_X$ model and the temperature dependent effective potential that enters 
in the analysis. Here we analyze the dynamics of the supercooled phase transition, incorporating both the percolation and the completion conditions, and emphasize the importance of using $R_*$ instead of $\beta$ for strongly supercooled transitions. In section~\ref{sec:xi} we examine the 
synchronous thermal evolution of the hidden and the visible sectors and its  influence on the analysis of the supercooled phase transition in the hidden sector, as well as the effect of
other cosmological constraints.  In section~\ref{sec:GW}, we present the resulting \gw power spectra, highlighting benchmark points that can fit the current PTA data. In section ~\ref{sec:ED} an analysis of the \gw energy density generated in the supercooled phase
transition is given.  
We summarize our conclusions and provide an outlook in section~\ref{sec:conclusion}.
Further mathematical details are given in the Appendix.

\section{Model \label{sec:model}}

\subsection{A hidden $U(1)_X$ model interacting with the visible sector }
The analysis of this work is based on the Lagrangian
$\mathcal{L}= \mathcal{L}_{\text{SM}}+ \Delta\mathcal{L}$ where 
$\mathcal{L}_{\text{SM}}$ is the standard model Lagrangian and $\Delta\mathcal{L}$
{is the Lagrangian for the hidden sector and its coupling to the visible sector}
\begin{align} \Delta\mathcal{L} = & -\frac{1}{4} F_{\mu\nu} F^{\mu\nu} -  \frac{\delta}{2} F_{\mu\nu} B^{\mu\nu}-|(\partial_\mu - i g_x A_\mu) \Phi|^2  \non  & + \bar{q} (i \gamma^\mu \partial_\mu - m_q) q  -  f_x \bar{q} \gamma^\mu q A_\mu - V_{0}^{\h}(\Phi), \label{bsm} \end{align}
 where
 
 {
\begin{align}
 V_{0}^{\h}&= -\mu_h^2\Phi\Phi^* +\lambda_h (\Phi^*\Phi)^2,~~
\Phi= \frac{1}{\sqrt 2} (\phi_c+ \phi+ i G^0_h).
\label{pot-hid}
\end{align}
}

Here $A_\mu$ is the gauge field associated with a hidden $U(1)_X$ symmetry, $\Phi$ is a complex scalar field { in the hidden sector and charged under 
$U(1)_X$}, $q$ is a dark fermion, and $B_\mu$ is the gauge field of the Standard Model's $U(1)_Y$ hypercharge symmetry. The field strength tensors are defined as $F_{\mu\nu} = \partial_\mu A_\nu - \partial_\nu A_\mu$ and $B_{\mu\nu} = \partial_\mu B_\nu - \partial_\nu B_\mu$. Thermal contributions to the zero-temperature effective potential {$V_{0}^{\h}(\Phi)$} enable a first-order phase transition, during which the scalar field $\Phi$ acquires a vacuum expectation value (VEV). This VEV generates masses for both the hidden sector gauge boson $A_\mu$ and the scalar field $\phi$ itself. Thus the effective  temperature dependent hidden sector potential including loop corrections 
is given by
{
\begin{align}
V^{\h}_{{\rm eff}}(\phi_c,T_h)&=V_{0}^{\h}(\phi_c)+
V_{\rm 1}^{\h}(\phi_c) 
+\Delta V_{T_h}^{\h}(\phi_c,T_h),
\label{pot-hid2}
\end{align}
}
where {$V^{\h}_1$} is the {zero temperature} one-loop 
Coleman-Weinberg potential and $\Delta V_{T_h}^{\h}$
is the finite thermal correction. Here for $V^{\h}_1$ we have
\begin{align}
    V_{\rm 1}^{\h}(\phi_c) = \sum_i \frac{g_i(-1)^{2s_i}}{64\pi^2}m_i^4(\phi_c)
  \left[\ln{\left(\frac{m_i^2(\phi_c)}{\Lambda^2}\right)}-{\cal C}_i\right],
\end{align}
where $g_i$ is the degrees of freedom of the particle and $s_i$ its spin
and $\mathcal{C}_i = \frac{5}{6}(\frac{3}{2})$ for gauge bosons(fermions). In this model we have three hidden sector fields $A_\mu,\phi,G_h^0$ contributing to effective potential and their field dependent masses  are 
 \begin{align}
  m_A^2(\phi_c)= g^2_x \phi_c^2,  ~~m_\phi^2(\phi_c)= -\mu_h^2+ 3 \lambda_h \phi_c^2,
  ~~m_{G^0_h}^2(\phi_c)= -\mu_h^2+ 3 \lambda_h \phi_c^2,
 \end{align}
The finite thermal correction $\Delta V_{T_h}^{\h}$ is given by
\begin{align}
    \Delta V_{T_h}^{\h}(\phi_c,T_h) &= \frac{T_h^4}{2\pi^2}\sum_{i=\text{bosons}}g_i\int_0^\infty dq~q^2\ln\left(1-\exp{\left(-\sqrt{q^2+\mathcal{M}_i^2(\phi_c,T_h)/T_h^2}\right)}\right)\nonumber\\
    &-\frac{T_h^4}{2\pi^2}\sum_{i=\text{fermions}}g_i\int_0^\infty dq~q^2\ln\left(1+\exp{\left(-\sqrt{q^2+\mathcal{M}_i^2(\phi_c,T_h)/T_h^2}\right)}\right), 
\end{align}
where $\mathcal{M}_i^2(\phi_c,T_h)$ are thermal corrected masses using Debye masses given by
\begin{align}
    \mathcal{M}_i^2(\phi_c,T_h) &= m_i^2(\phi_c) +\Pi_i(T_h)\\
    \Pi_{A}(T_h) &=  \frac{2}{3}g_x^2 T_h^2\\
    \Pi_{\phi}(T_h) &= \frac{1}{4}g_x^2 T_h^2 + \frac{1}{3}\lambda_hT_h^2.
\end{align}
Calculational details for the above can be found in \cite{Feng:2024pab}.
{ 
We pause here to remark on the issue of gauge invariance and effective potential. Although the effective potential is gauge invariant at its extrema values, in general it is gauge dependent \cite{Jackiw:1974cv} at any 
$T$ including $T=0$. The effective potential can be made gauge invariant by inclusion of  correction terms 
 using the Nielsen identity\cite{Nielsen:1975fs,Fukuda:1975di}.
However, the Nielsen indentity is typically implemented in loop order~\cite{Metaxas:1995ab,Patel:2011th,Andreassen:2014eha,Lofgren:2021ogg}
but the complete inclusion of gauge invariance in effective potentials would likely involve use of lattice gauge analysis\cite{Munehisa:1984is} especially in the strong coupling regime and in analyses involving confinement.
 In work here we have used the Landau gauge and gauge dependent corrections are neglected.
 The sensitivity of the effective potential
 due to variations in $\xi^{\rm gauge}$ is discussed in 
 later part of the paper.
 }
Returning to our analysis,
according to Eqs.~(\ref{bsm},\ref{pot-hid}), one finds that 6 parameters define the model: $g_x, m_q, \delta, f_x, \mu_h$ and $\lambda_h$.
An additional important parameter is  $\xi_0$, which is the initial temperature 
ratio of the hidden sector temperature and the temperature of the visible sector
at the high temperature,
a topic  which we will discuss in further detail in Section \ref{sec:3}. 
\subsection{Supercooled Hidden Sector Phase Transition from Hidden $U(1)_X$ Model\label{sec:U1supercooled}}

As discussed in the introduction, supercooled phase transitions have attracted attention as
 they can yield a sufficiently small transition rate $\beta_*/H_*$ (here * represents the time when the \gws are generated, which is {taken} to be the percolation temperature in this work), compatible with PTA observations. However, for strongly supercooled phase transitions, using $\beta_*$ to characterize the transition rate is not valid and leads to a high 
 {degree of} error because it is just a linear coefficient in {a} Taylor expansion \cite{Athron:2023xlk,Megevand:2016lpr}. 
The mean bubble separation $R_*$ was first introduced in \gws fits {where for} non-supercooled phase transitions, studies have shown that $R_*$ can be well approximated by \cite{Enqvist:1991xw}:
\begin{align}
    R_{*}\approx (8\pi)^{1/3}\frac{v_w}{\beta_*}\label{eq:approx}, 
\end{align}
where $v_w$ is the bubble wall velocity. However, this approximation breaks down for supercooled phase transitions. In such cases, $R_*$ should be determined directly from its definition for { \gw analysis} rather than derived from $\beta_*$, as established in several studies \cite{Athron:2023mer, Athron:2023rfq, Caprini:2019egz}.

 In this section, we discuss the dependence of $H_*R_*$ on the 
  {parameters $\lambda_h,g_x$}, 
 while the previous studies have focused on $\beta_*/H_*$ \cite{Breitbach:2018ddu,Bringmann:2023iuz,Banik:2024zwj}.
In the computation of $R_*$ and $\beta_*$, we need to first determine when the phase transition occurs and
 the \gws are generated. As discussed in \cite{Athron:2022mmm,Athron:2023mer, Caprini:2019egz, Wang:2020jrd}, the percolation temperature $T_{h,p}$ appears as  
  the appropriate temperature for the production of \gws during a supercooled phase transition. Here, $T_{h,p}$ denotes the temperature of the hidden sector at which 71$\%$ of the universe remains in the false vacuum, and is defined by 
\begin{align} 
P_f(T_{h,p}) &= 0.71, \\
         P_f(T_h)&=\exp\left(-\frac{4\pi}{3}v_w^3\int_{T_h}^{T_{h,c}}dT'\frac{\Gamma(T_h')}{T_h'H^4(T_h')}\left(\int_{T_h}^{T_h'}dT_h''\frac{1}{H(T_h'')}\right)^3\right),\label{eq:percolation} \\
         \Gamma(T) &\simeq T^4 \left(\frac{S(T)}{2\pi}\right)^{3/2}\exp{(-S(T))},
         \label{pf}
     \end{align}
where $S(T) = S_3(T)/T$ with $S_3$ being the three-dimensional Euclidean action (bounce action), $T_{h,c}$ is the critical temperature and $H$ is the hubble parameter. Here, we use \textbf{CosmoTransitions} \cite{Wainwright:2011kj} to numerically find Euclidean action $S_3(T)$ with {the} given effective potential. In this subsection, we assume the temperature ratio between the hidden and the visible sector $\xi = \frac{T_h}{T_v} = 1$ and $v_w = 1$ for a trivial discussion. We will discuss the non-trivial case later in section $\ref{sec:3.3}$. Upon determining $T_{h,p}$, the parameter $\beta_*/H_*$ can be calculated {using}

\begin{align} \frac{\beta_*}{H_*} = T_{h} \left. \frac{dS(T_h)}{dT_h} \right|_{T_h = T_{h,p}}. \label{eq:betaoverH}
\end{align} 
 To achieve a small $\beta/H$, the derivative $\frac{dS(T_h)}{dT_h}$ should approach zero near $T_h = T_{h,p}$. For $R_*$, we calculate it directly from the bubble number density 
 so that
\begin{align}
    R_*(T_h) = (n_B(T_h))^{-\frac{1}{3}} = \left(T_h^3\int_{T_h}^{T_{h,c}}dT_h'\frac{\Gamma(T_h')P_f(T_h')}{{T_h'}^4H(T_h')}\right)^{-\frac{1}{3}}.\label{eq:Rstar}
\end{align}
\begin{figure}[h] 
\centering 
\includegraphics[width=0.75\linewidth]{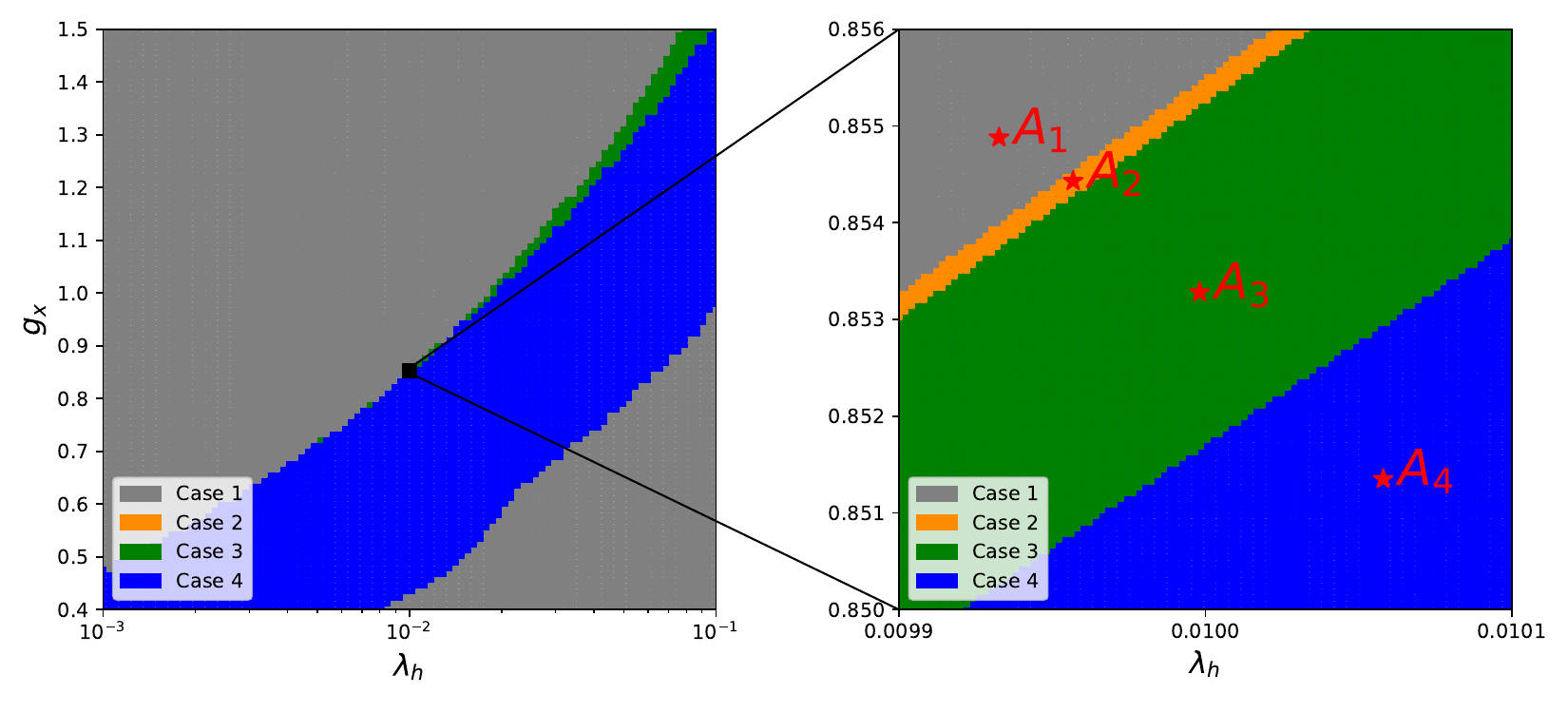} 
\caption{Left panel: The parameter space in the $\lambda_h$-$g_x$ plane divided into four regions with distinct colors, each corresponding to a different phase transition case for the
cases (Case 1 -Case 4) discussed in section (\ref{sec:U1supercooled}) where Case 2 occupies a region too small to be visible. Right panel:  A magnified view of a small region of the left panel is displayed. Here four characteristic points labeled $A_1-A_4$ defined in {Table \ref{tab:A1A4}}
 are marked on the plot. 
}
\label{fig:parameter-space} 
\end{figure}
\begin{figure}[h] 
\centering  
\includegraphics[width=0.50\linewidth]{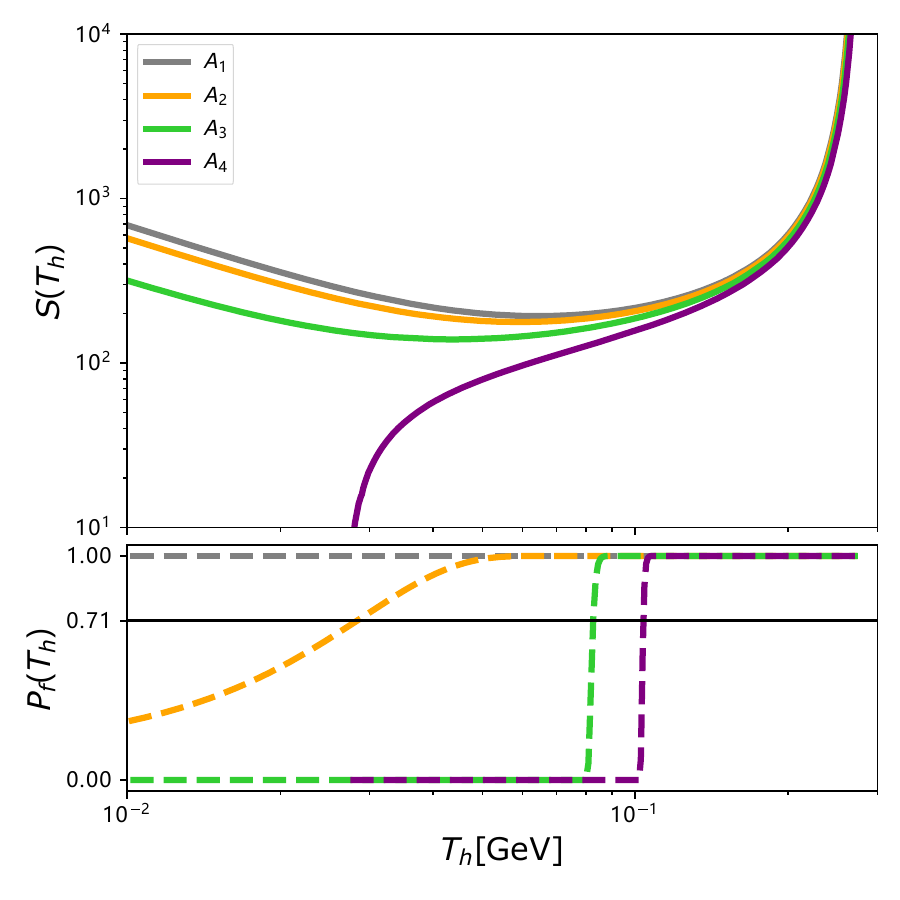} 
\caption{The bounce action $S(T_h)$ (solid lines) and 
$P_f(T_h)$ (dashed lines) {defined by Eq.(\ref{pf}) where $(1-P_f(T_h))$ is the true vacuum fraction,} 
are plotted vs $T_h$ for four characteristic points labeled $A_1-A_4$ defined by $g_x$ and $\lambda_h$.}
\label{fig:parameter-space1} 
\end{figure}

The {left} panel of Fig.~\ref{fig:parameter-space} shows the parameter space spanning $\lambda_h \in (10^{-3}, 2 \times 10^{-1})$ versus $g_x \in (0.4, 1.5)$, with the {right panel}
showing the region $\lambda_h \in (0.0099, 0.0101)$ versus $g_x \in (0.850, 0.856)$. The space is partitioned into four distinct regions, each colored differently to represent different types of phase transitions, which are:

    \noindent{\bf Case 1:}
   Here  first-order phase transitions do not occur. There are two such regions 
     in the {left panel of Fig.~\ref{fig:parameter-space}. The first one is   
     the upper dark region and is called the ``ultracooled" region, where the potential barrier becomes too large for the transition to achieve percolation. The second one is the lower dark } ``crossover" region, where the transition from the false vacuum to the true vacuum is smooth and continuous, with no barriers between phases, thus precluding a first-order phase transition. The supercooled first-order phase transition of interest lies on the boundary of the ``ultracooled" region.
    
        \noindent{\bf Case 2:}    
  Here first-order phase transitions either do not complete or the physical volume of the false vacuum does not decrease at $T_{h,p}$. In this case, while percolation is successfully achieved with $P_f = 0.71$, the phase transition either fails to complete 
    ({$P_f\geq  0.01$} at the end)  
  or the physical volume of the false vacuum $V_{false} \propto a^3(t)P_f(t)$ increases due to a growing scale factor $a(t)$ via Hubble expansion, preventing bubble collisions and thus \gw generation. Traditionally, the constraint on a decreasing volume of the false vacuum is satisfied by \cite{Freese:2022qrl,Turner:1992tz}:
    \begin{align}
        \frac{\beta_*}{H_*} > 3. \label{eq:betaoverHlarger3}
    \end{align}
    However, since using $\beta_*$ for a supercooled phase transition leads to a significant error, an alternative constraint is employed \cite{Athron:2022mmm, Athron:2023mer}:
    \begin{align}
      3 + \left.T_{h,p}\frac{d\mathcal{V}_t^{ext}}{dT_h}\right|_{T_h = T_{h,p}} < 0,  
    \end{align}
    where $-\mathcal{V}_t^{ext}$ is the exponent in Eq.~ (\ref{eq:percolation}).

        \noindent{\bf Case 3:}    The first-order phase transition successfully completes with a permanent potential barrier (i.e., the barrier persists even at $T_h = 0$). A typical feature of this case is that the action curve $S(T)$ has a U-shape. With such a U-shape, we can achieve a small transition rate since $\frac{dS(T_h)}{dT_h}$ can approach zero. This case is expected to have a sufficiently small transition rate with strong enough transition strength to fit PTA signals.
        
        \noindent{\bf Case 4:}    
   The  first-order phase transition successfully completes with the metastable false vacuum disappearing at some non-zero temperature. A typical feature of this case is that the action curve $S(T)$ has a ``tangent-like" shape. In this case, the transition rate is generally large because the action curve is a monotonically increasing function, meaning $\frac{dS(T_h)}{dT_h}$ is typically large. This case is expected to exhibit either a weak supercooled phase transition or a non-supercooled phase transition.

\begin{table}[h]
 \centering
 \small
     \begin{tabular}{clclcl}
   \hline
Type&Point& $g_x$ &$\lambda_h$    \\
\hline
Case 1 & $A_1$ & 0.8549&0.009933   \\
\hline
Case 2 & $A_2$ & 0.8544&0.009957   \\
\hline
Case 3 & $A_3$ & 0.8533&0.009998   \\
\hline
Case 4 & $A_4$ & 0.8514&0.010058   \\
\hline
   \end{tabular}
 \caption{   }
  \label{tab:A1A4}
\end{table}
Four model model points labeled $A_i$ ($i = 1,2,3,4$) one for each of the  Cases 1-Case 4 
defined in {Table \ref{tab:A1A4} }
are shown in Fig.~\ref{fig:parameter-space1} where the bounce action $S$ and the  percolation probability $P_f$ versus $T_h$ are plotted in Fig.~\ref{fig:parameter-space1}  to illustrate each case. 
For $A_1$, the percolation probability $P_f(T_h)$ is a straight horizontal line equal to 1, indicating that percolation never occurs. For $A_2$, the transition is so slow that $P_f$ takes considerable time to decrease from 1 to 0, and consequently, we find the physical volume of the false vacuum increasing at the percolation temperature. For $A_3$, the transition is fast and generates {gravitational waves}. In this case, we observe a U-shaped action curve. For $A_4$, the false vacuum disappears in finite time, ensuring the completion of the transition. It exhibits a typical ``tangent-like" action curve  characteristic of weakly supercooled or non-supercooled phase transitions. From the right panel of Fig.~\ref{fig:parameter-space}, we find that by decreasing $g_x$ or increasing $\lambda_h$, phase transitions progress from Case 1 to Case 4. Case 3 is of particular interest because it generates gravitational waves strong enough to fit PTA signals. From the left panel of Fig.~\ref{fig:parameter-space}, we observe that Case 3 occupies a larger area as $\lambda_h$ increases, indicating that a strongly supercooled phase transition favors a larger $\lambda_h$. Conversely, when $\lambda_h$ is very small, Case 3 may disappear entirely, which suggests that a supercooled phase transition is prohibited.

\begin{figure}[h] 
\centering 
\includegraphics[width=0.45\linewidth]{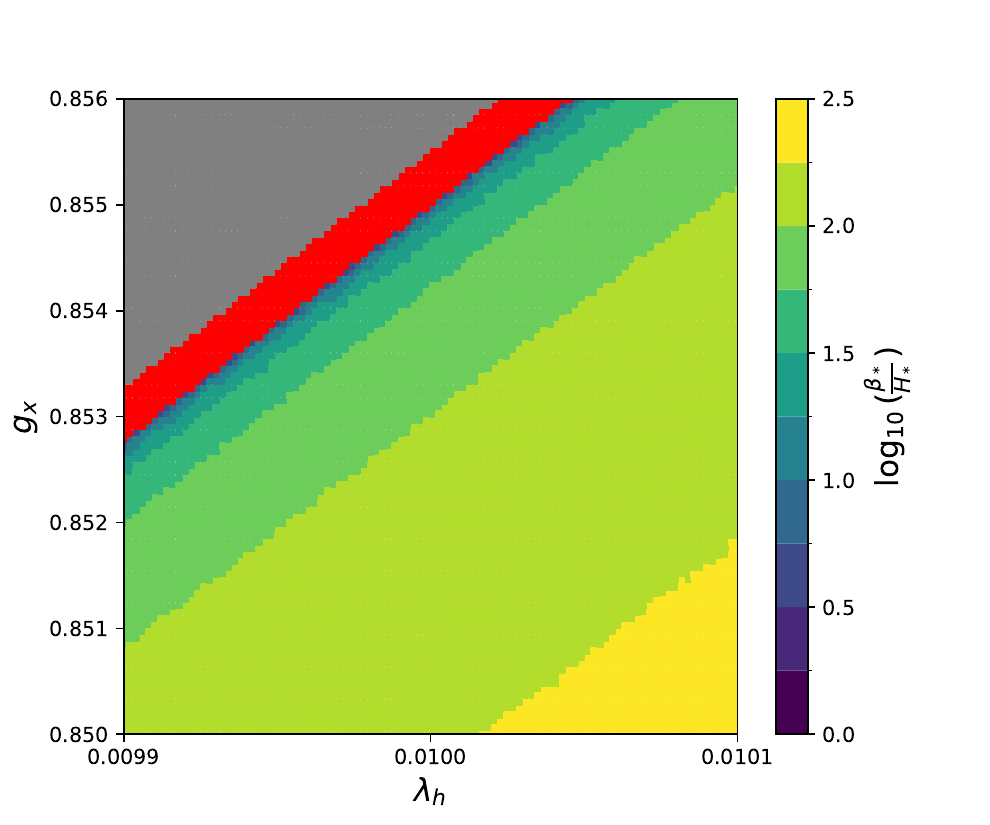} 
\includegraphics[width=0.45\linewidth]{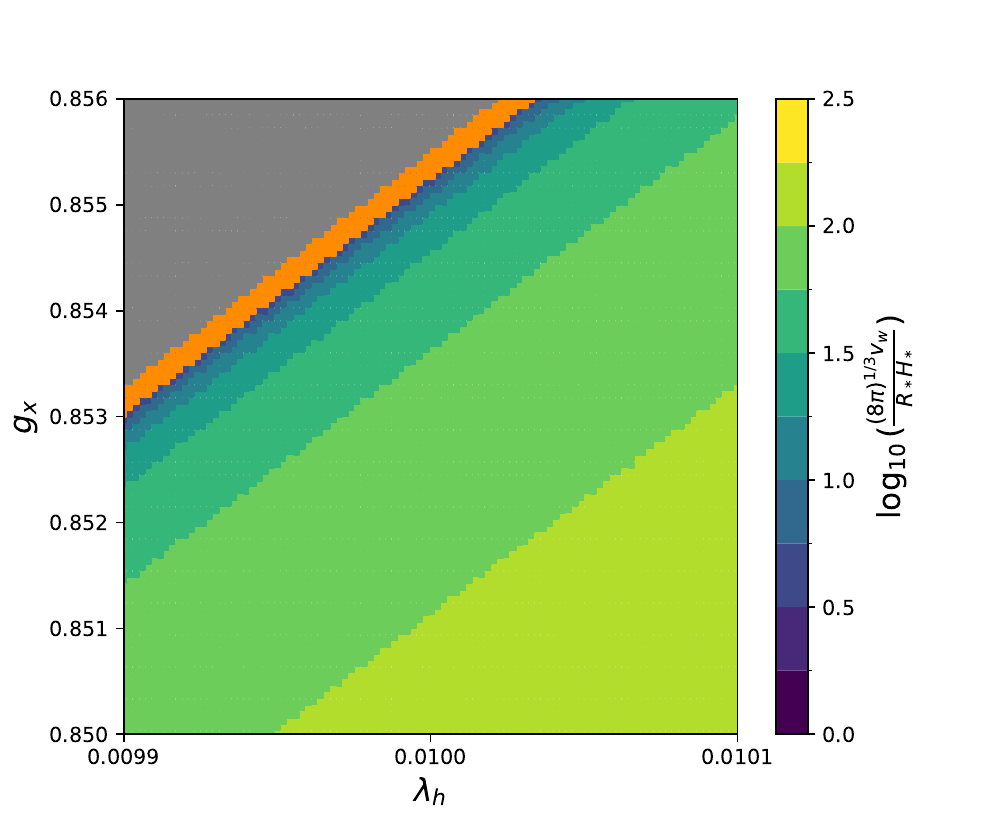} 
\caption{Left panel: An display of {log$_{10}(\beta_*/H_*)$} given by Eq.(\ref{eq:betaoverH}) in the $\lambda_h$-$g_x$ plane.
The gray region represents Case 1. The red region shows the constraint given by $\beta_*/H_*>3$. The remaining region shows a color map of $\beta_*/H_*$ values.
Right panel: The gray and orange regions represent Case 1 and Case 2, respectively. The remaining region shows a color map of $\frac{(8\pi)^{1/3}v_w}{R_*H_*}$  by Eq.(\ref{eq:Rstar}). }
\label{fig:Randbeta} 
\end{figure}
To demonstrate the significant error introduced by the approximation in Eq.~(\ref{eq:approx}), we analyze two terms: $\beta_*/H_*$ and $\frac{(8\pi)^{1/3}v_w}{R_*H_*}$. These terms should have comparable values if the approximation holds. In Fig.~\ref{fig:Randbeta}, we present two color maps of these terms within the parameter space $\lambda_h \in (0.0099, 0.0101)$ versus $g_x \in (0.850, 0.856)$, which correspond to the zoomed-in region shown in Fig.~\ref{fig:parameter-space}. {In the plots of Fig.~\ref{fig:Randbeta}}, the gray and the orange regions indicate Case 1 and Case 2, respectively. The red region represents the constraint imposed by $\beta_*/H_*>3$, which is commonly considered when using $\beta_*/H_*$. 
The color maps of the {left panel and the right panel of Fig.~\ref{fig:Randbeta}}
 exhibit stark differences, clearly demonstrating that the approximation in Eq.~(\ref{eq:approx}) fails when analyzing supercooled phase transitions. Further, we observe that for given values of $\lambda_h$ and $g_x$, $\beta_*/H_*$ consistently yields larger values than $\frac{(8\pi)^{1/3}v_w}{R_*H_*}$, indicating that using $\beta_*/H_*$ leads to an underestimation of the final \gw power spectrum. Notably, the red region substantially exceeds the orange region, suggesting that applying Eq.~(\ref{eq:betaoverHlarger3}) incorrectly excludes physically viable parameters. These parameters lie near the boundary of the ``ultracooled'' region, indicating {that} strongly supercooled phase transitions 
 {can arise in this region that can potentially generate \gw power spectrum} powerful enough to match PTA signals.
 
  \section{Why thermal history is important for a
supercooled phase transition\label{sec:xi}}
The temperature ratio $\xi$  between the hidden and the visible sectors, enters 
in hidden sector phase transitions, as discussed in several previous studies \cite{Breitbach:2018ddu,Bringmann:2023opz,Bai:2021ibt, Li:2023bxy,Feng:2024pab}.
 The effective number of relativistic species, {$N_{\rm eff}$}
  and the transition strength, $\alpha_{\rm tot}$, are two factors that are significantly influenced by the temperature ratio $\xi$. Generally, there are two types of transition strength parameters, as discussed in \cite{Bringmann:2023opz,Bai:2021ibt}:
\begin{align}
\label{alphatot}
    \alpha_{\rm tot} &= \frac{\Delta \Bar{\theta}(T_{h,p})}{\rho_{\rm rad}^v(T_{v,p}) + \rho_{\rm rad}^h(T_{h,p})}, \\
    \alpha_h &= \frac{\Delta \Bar{\theta}(T_{h,p})}{\rho_{\rm rad}^h(T_{h,p})}, \label{Eq.alphastar}
\end{align}
where $T_{h,p}$ and $T_{v,p}$ are the percolation temperatures in the hidden and visible sectors, respectively. {Below we show that $\alpha_{tot}$ which controls the \gw power spectrum is proportional to $\xi_p$ where $\xi_p = T_{h,p}/T_{v,p}$ indicating the strong role
that $\xi$ plays in generating \gws strong enough to match the  PTA signal.}
In Eq.(\ref{Eq.alphastar}), $\Delta \Bar{\theta}$ represents the amount of vacuum energy released during the transition \cite{Giese:2020znk,Giese:2020rtr}, and is defined as:
\begin{align}
    \Delta \Bar{\theta}(T_h)& = \Bar{\theta}_f(T_h) - \Bar{\theta}_t(T_h),\\
     \Bar{\theta}_i(T_h) &= \frac{1}{4}\left(\rho^h_i - \frac{p^h_i(T_h)}{c_{s,t}^2(T_h)}\right),
\end{align}
where
\begin{align}
    p^h_i(\phi_i,T_h) &=  \frac{\pi^2}{90}g^h_{eff}(T_h)T_h^4 - V^h_{eff}(\phi_i,T_h),\label{eq:phi}\\
    \rho^h_i(\phi_i,T_h) &= T_h\frac{\partial p^h_i}{\partial T_h} - p^h_i,\label{eq:rhohi}\\
    c^2_{s,t}(T_h) &= \frac{\partial p^h_t}{\partial \rho^h_t} = \frac{\partial p^h_t/\partial T_h}{\partial \rho^h_t/\partial T_h},
\end{align}
Here the notation ``$f$" and ``$t$" stand for the  false vacuum and the true vacuum.
{As noted above} the parameter $\alpha_{tot}$ enters directly in the \gw  power spectrum analysis while 
 $\alpha_h$ enters in hydrodynamic computations for the hidden sector to determine the efficiency factor $\kappa$ and the bubble wall velocity $v_w$ when the visible sector is either very weakly coupled to or decoupled from the hidden sector. However,  if there is a strong coupling between the visible and hidden sectors, the situation can be quite different. In this work, we assume that the visible sector is weakly coupled to the hidden sector during the phase transition, so $\alpha_h$ is primarily used for energy budget calculations. In the case of a supercooled phase transition, {we expect} $\alpha_h \gg 1$. However, to measure the strength of the first-order phase transition and the \gws it generates, $\alpha_{\rm tot}$ should be used instead.

{In the class of models discussed here}, we typically have $g_{\rm eff}^v \gg g_{\rm eff}^h$,
where  $g_{\rm eff}^v$ is the number of degrees of freedom in the visible sector and
 $g_{\rm eff}^h$ is the number of degrees of freedom in the hidden sector. 
 In this case, we expect that the main contribution to the radiation energy density comes from the visible sector, where the radiation energy density is given by
$
   \rho_{\rm rad,v}(T_h) = \frac{\pi^2}{30}g_{\rm eff}^v\left(\frac{T_{h,p}}{\xi_p}\right)^4
$.
Consequently, we find that \cite{Breitbach:2018ddu}
\begin{align}
   \alpha_{\rm tot}&\sim\frac{\Delta \Bar{\theta}(T_{h,p})}{\rho_{\rm rad}^v(T_{v,p}) } \sim\frac{\Delta \Bar{\theta}(T_{h,p})}{\frac{\pi^2}{30}g_{\rm eff}^v\left(\frac{T_{h,p}}{\xi_p}\right)^4 }\propto \xi_p^4,
\end{align}
{which means it is possible to have $\alpha_{tot}\ll 1$ even for a strong supercooled phase transition with $\alpha_h\gg 1$.} To generate observable signals in PTAs, it is desirable to maximize $\alpha_{\rm tot}$, which in turn means maximizing $\xi_p$.  On the other hand, the effective number of additional neutrino species, 
\begin{align}
    \Delta N_{\rm eff} = \frac{4}{7}g^h_{eff}(T_{h,BBN})\left(\frac{11}{4}\right)^{4/3}\xi_{BBN}^4\label{eq:Neff}, 
\end{align}
scales as $\xi_{BBN}^4$, and observations from BBN and CMB impose the constraint $\Delta N_{\rm eff} < 0.3$ \cite{Planck:2018vyg}. These two conditions, i.e., maximizing $\alpha_{\rm tot}$ and limiting $\Delta N_{\rm eff}$, are in conflict, which substantially constrains the viable parameter space for $\xi$.

This conflict becomes more pronounced for a supercooled phase transition when reheating becomes non-negligible. To find the reheating temperature $T_{h,reh}$, we assume that reheating completes instantaneously around the time of bubble percolation. Energy conservation (see Section~\ref{sec:ED}) then gives us:
\begin{align}
   \rho^h_{rad}(T_{h,reh}) \simeq \rho^h_{rad}(T_{h,p}) + \rho^h_{vac}(T_{h,p}) \label{Eq.reheat}
\end{align}
For strong phase transitions, such as those which are of interest in this work, we have $\rho_{\rm vac}^h(T_h) \simeq \Delta \Bar{\theta}(T_h)$. Combining Eqs. (\ref{Eq.reheat}) and (\ref{Eq.alphastar}), we obtain:
\begin{align}
   T_{h,reh} = \left(1+\alpha_h\right)^{\frac{1}{4}}\left(\frac{g^h_{eff}(T_{h,p})}{g^h_{eff}(T_{h,reh})}\right)^{\frac{1}{4}}T_{h,p}
\end{align}
Solving this equation yields $T_{h,reh}$. Since reheating occurs so rapidly that all released heat is transferred to the hidden sector, the temperature of the visible sector remains unchanged:
\begin{align}
   T_{v,reh} = T_{v,p}
\end{align}
This leads to a new temperature ratio after reheating:
\begin{align}
   \xi_{reh} = \frac{T_{h,reh}}{T_{v,reh}}
\end{align}
In a supercooled phase transition, we know that $\alpha_h\gg 1$, which implies $T_{h,reh}\gg T_{h,p}$ and consequently $\xi_{reh}\gg \xi_p$. This exacerbates the conflict since we need $\xi_{reh}$ to be as small as possible to satisfy the BBN constraint on $\Delta N_{\rm eff}$, while simultaneously requiring $\xi_p$ to be as large as possible to generate 
\gws
strong enough to match PTA signals. A detailed analysis by \cite{Bringmann:2023opz} shows that for a stable hidden sector, the viable parameter space is severely restricted. However, as also noted in \cite{Bringmann:2023opz}, a dark sector decaying at pre-BBN temperature could resolve this conflict. We will examine this scenario with a detailed analysis in the following subsection.\\

\subsection{Thermal evolution of coupled hidden and visible sectors\label{sec:3}}
 Thermal evolution of the hidden and the visible sectors are affected by coupling between
 the two sectors. Such couplings can exist via a variety of portals which include
 kinetic mixing of gauge fields \cite{Holdom:1985ag}, via Higgs coupling to the visible and the hidden sectors \cite{Patt:2006fw} and via Stueckelberg mass mixing\cite{Kors:2004dx,Feldman:2007wj,Du:2022fqv}
 among others. In this work, we assume  kinetic mixing for convenience. We generally use the notation $T_h$ for the hidden sector temperature and $T_v$ for the visible sector temperature. To describe the synchronous evolution of the hidden and the visible sectors we define
\begin{align}
    T_h=\xi(T_v) T_v \label{eqxi}
\end{align} 
where $\xi(T_v)$ is an evolution function that relates the visible and the hidden sector
temperatures.  
It is also convenient to sometime use $\zeta(T_h)$ where 
\begin{align}
    T_v=\zeta(T_h) T_h 
    \label{eqzeta}
\end{align} 
which allows to fix the temperature in the visible sector given the temperature in the hidden sector.{ $\zeta(T_h)$ and $\xi(T_v)$ are simply related: $\zeta(T_h)= \xi^{-1}(T_v)$ and
$\xi(T_v)$ and $\zeta(T_h)$ are alternately used as needed to simplify notation.}
To find $\xi(T_v)$, we start from the coupled Boltzmann equations in an expanding universe so that
\begin{align}
&\frac{d\rho_v}{dt} + 3 H(\rho_v+p_v)=j_v,\non
&\frac{d\rho_h}{dt} + 3 H(\rho_h+p_h)=j_h.
\end{align}
Here $\rho_v, p_v$ and $\rho_h, p_h$ are the energy and pressure densities
for the visible and hidden sectors, 
  and where $j_v,j_h$ encode in them all the possible processes exchanging energy between these sectors~\cite{Aboubrahim:2021ycj,Aboubrahim:2022bzk}.
The evolution equation for $\xi(T)$ is given by
 \cite{Aboubrahim:2020lnr,Li:2023nez,Nath:2024mgr}
\begin{align}
\frac{d\xi}{dT_v}= \left[ -\xi \frac{d\rho_h}{dT_h} +
\frac{4H\sigma_h\rho_h-j_h}{4H\sigma\rho-4H\sigma_h\rho_h+ j_h} \frac{d\rho_v}{dT_v}\right] (T \frac{d\rho_h}{dT_h})^{-1}.
\label{DE1}
\end{align}
where $\sigma = \frac{3}{4}(1+\frac{p_v+p_h}{\rho_v+\rho_h}), \sigma_h = \frac{3}{4}(1+\frac{p_h}{\rho_h})$.
In the analysis using the above one needs to pay attention to the limit when 
$\rho_h\rightarrow 0$ which happens for the case of a decaying dark sector.
 To study the thermal evolution of a decaying hidden sector, we account for the energy density not only through thermal equilibrium analysis but also by including the contribution from the relic abundance. Details on this can be found in Appendix E of \cite{Feng:2024pab}. 
 In Eq.({\ref{DE1})  $j_h$  depends {on the particle content and their interactions in the
 hidden sector}, which can be gotten by solving the Boltzmann equations for the yields discussed below:

\begin{align}
        \frac{dY_q}{dT}  =& -\frac{\bold{\mathbb{s}}}{H}\left( \frac{d\rho_v/dT}{{4\sigma\rho-4\sigma_h\rho_h}+j_h/H}\right) \Big[\frac{1}{2}\left<\sigma v\right>_{q\bar{q} \rightarrow i\bar{i} }(T)(Y_q^{eq}(T)^2-Y_q^2) \non
   &  -\frac{1}{2}\left<\sigma v\right>_{q\bar{q} \rightarrow \gamma'{\gamma'} }(T_h)\left(Y_q^2-Y_q^{eq}(T_h)^2\frac{Y_{\gamma'}^2}{Y_{\gamma'}^{eq}(T_h)^2}\right)
   \Big]\,,\label{DE2}
  \end{align}
  
  \begin{align}
    \frac{dY_{\gamma'}}{dT}  =& -\frac{\bold{\mathbb{s}}}{H}\left( \frac{d\rho_v/dT}{{4\sigma\rho-4\sigma_h\rho_h}+j_h/H}\right)  \left[ \frac{1}{2}\left<\sigma v\right>_{q\bar{q} \rightarrow \gamma'{\gamma'} }(T_h)\left(Y_q^2-Y_q^{eq}(T_h)^2\frac{Y_{\gamma'}^2}{Y_{\gamma'}^{eq}(T_h)^2}\right) +\right.\non
    &\frac{1}{2}\left<\sigma v\right>_{\phi\bar{\phi} \rightarrow \gamma'{\gamma'} }(T_h)\left(Y_\phi^2-Y_\phi^{eq}(T_h)^2\frac{Y_{\gamma'}^2}{Y_{\gamma'}^{eq}(T_h)^2}\right)+\left<\sigma v\right>_{i\bar{i}\rightarrow \gamma' }(T)Y_{i}^{eq}(T)^2\non
    &\left.-\frac{1}{\bold{\mathbb{s}}}\left<\Gamma_{\gamma'\rightarrow i\bar{i}}(T_h)\right>Y_{\gamma'}+\left<\Gamma_{\phi\rightarrow \gamma'\gamma'}(T_h)\right>\left(Y_{\phi}-Y^{eq}_\phi(T_h)\frac{Y_{\gamma'}^2}{Y^{eq}_{\gamma'}(T_h)^2}\right)\right]\,,
    \label{DE3}
    \end{align}
  
    \begin{align}
      \frac{dY_\phi}{dT}  =& -\frac{\bold{\mathbb{s}}}{H}\left( \frac{d\rho_v/dT}{{4\sigma\rho-4\sigma_h\rho_h}+j_h/H}\right)  \big[-\frac{1}{2}\left<\sigma v\right>_{\phi\bar{\phi} \rightarrow \gamma'{\gamma'} }(T_h)\left(Y_\phi^2-Y_\phi^{eq}(T_h)^2\frac{Y_{\gamma'}^2}{Y_{\gamma'}^{eq}(T_h)^2}\right) \non
    &-\frac{1}{\bold{\mathbb{s}}}\left<\Gamma_{\phi\rightarrow \gamma'\gamma'}(T_h)\right>\left(Y_{\phi}-Y^{eq}_\phi(T_h)\frac{Y_{\gamma'}^2}{Y^{eq}_{\gamma'}(T_h)^2}\right) \big]\,.\label{DE4}
\end{align}
 {In the set of equations above $\bold{\mathbb{s}}$ is the entropy density and $Y_i$ are 
 the yields defined by $Y_i = n_i/\bold{\mathbb{s}}$}
To solve the differential equation set, Eq.~(\ref{DE1},\ref{DE2},\ref{DE3},\ref{DE4}), we need to set up the initial values for $\xi$ and the yields $Y_{\gamma'}, Y_q$, and $Y_\phi$. For a freeze-in process {the initial value of $\xi$ is labeled $\xi_0$ and is taken to be zero.
More generally we will assume $\xi_0$ to be a free parameter.} More details may be found in Section 2.2 of \cite{Feng:2024pab}.

\begin{figure}[h]
    \centering
    \includegraphics[width=0.32\linewidth]{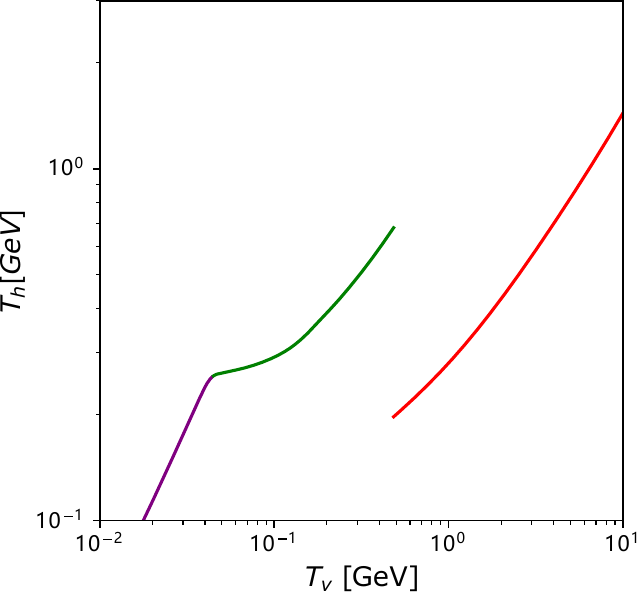}
    \includegraphics[width=0.32\linewidth]{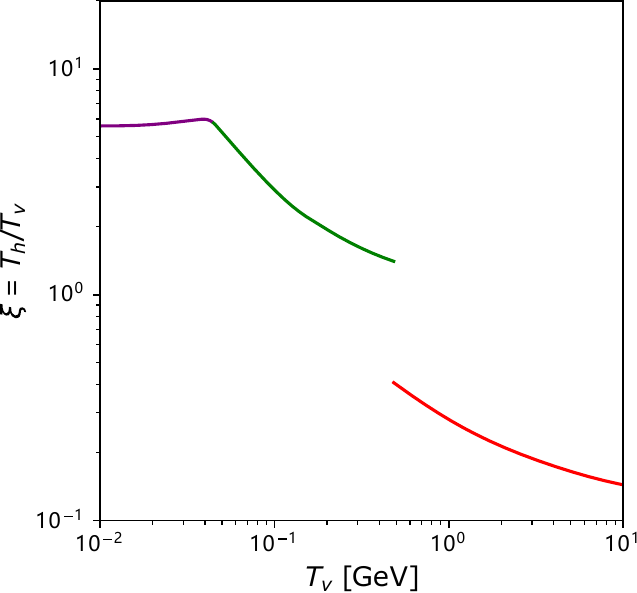}
    \includegraphics[width=0.32\linewidth]{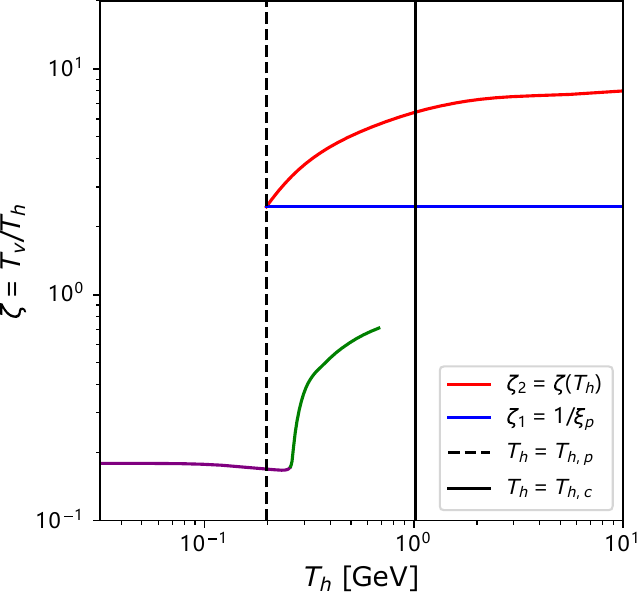}
    \caption{Temperature evolution of the hidden and visible sectors for benchmark point BP1 from Table \ref{tab:benchmarks2}. The evolution shows three distinct epochs: Phase 1 (red), before reheating; Phase 2 (green), after the reheating; and Phase 3 (purple), 
 after the hidden sector particles decouple from each other.
    Left panel: Evolution of $T_h$ v.s. $T_v$. Middle panel: Evolution of $\xi$ v.s. $T_v$. Right panel: Evolution of $\zeta$ v.s. $T_h$. The vertical dashed line marks th
      $T_h=T_{h,p}$ and the solid line marks the critical temperature for the hidden sector $T_h = T_{h,c}$. The blue line is the case $\zeta(T_h) = 1/\xi_p = \text{const}$. }
    \label{fig:newxievo1}
\end{figure}

The temperature evolution of the hidden and visible sectors for model BP1 is analyzed in Fig.(\ref{fig:newxievo1}). This evolution can be divided into three distinct epochs: Phase 1 (red), occurring before the phase transition and reheating; Phase 2 (green), following the phase transition and reheating; and Phase 3 (purple), when the hidden sector particles decouple from each other. 
During Phases 1 and 2, the evolution of $\xi(T_v)$ is governed by Eq.(\ref{DE1}). There is a discontinuous jump between Phase 1 and Phase 2, corresponding to the instantaneous reheating (analyzed in the previous section).

According to Table~\ref{tab:benchmarks2}, for BP1 the lightest hidden sector particle (LHP) is $\phi$ with a mass of $1.28$ GeV. When the hidden sector temperature falls below this mass, the sector becomes non-relativistic, and its particles are expected to annihilate and decay into other particles. However, if the coupling between the hidden and visible sectors is very small (as is the case in this work), the radiational energy of the hidden sector converts into chemical potential rather than transferring to the visible sector, yielding the entropy density:
\begin{align}
    s_h \simeq \frac{m_\phi-\mu_\phi}{T_h}n_\phi.
\end{align}

If the hidden sector maintains efficient number-changing processes, such as $\phi\phi\phi\to\phi\phi$, the chemical potential vanishes ($\mu_\phi=0$) while preserving the hidden-sector entropy. In this scenario, the hidden sector effectively consumes its own particles to maintain its temperature, causing $\xi$ to increase continuously ($\zeta$ decreasing). This mechanism is known as {``cannibalism."} For detailed discussions of cannibalism and the associated phase transition, see Refs.~\cite{Farina:2016llk,Ertas:2021xeh,Bringmann:2023opz,Bringmann:2023iuz}.
The  cannibalism persists after the phase transition concludes, continuing into Phase 2. The process terminates when either the number-changing processes become inactive or the hidden sector loses thermal contact and particles decouple from each other.  
In the analysis of this work, the latter case stops the cannibalism—specifically, when the interaction rate between hidden sector particles becomes smaller than the Hubble parameter. At this point, the hidden sector particles freeze out, marking the transition to Phase 3.

During Phase 3, changes in the hidden sector temperature can occur only due to the universe's expansion, similar to the visible sector. Consequently, the visible and the hidden sector temperatures evolve at the same rate, resulting in $\xi\sim\zeta\sim\text{const}$ during this phase.
Once the cannibalism phase ends, the hidden sector becomes fully non-relativistic, causing $g^{h,rad}_{eff}$ to decrease exponentially. This leads $\Delta N_{eff}$ to rapidly approach zero, indicating that the hidden sector particles' contribution to the radiation energy density becomes negligible. Consequently, the constraint on $\Delta N_{eff}$ is naturally satisfied. Meanwhile, the middle panel demonstrates that $\xi$ can reach $\mathcal{O}(1)$ at phase transition without violating the $\Delta N_{\rm eff}$ constraint. Thus, in the case of a decaying hidden sector, the apparent tension between $\alpha_{tot}$ and $\Delta N_{\rm eff}$ is resolved.

On the right panel, we indicate both $T_{h,p}$ and $T_{h,c}$ on the plot. The graph shows that the temperature ratio changes dramatically between $T_{h,c}$ and $T_{h,p}$. 
{In most previous works in the literature, the} temperature ratio was generally assumed to be constant, as indicated by the {horizontal}  blue line. In the next section, we will demonstrate how $\zeta(T_h)$ is involved in analyzing a hidden supercooled phase transition, highlighting the necessity of considering synchronous evolution {hidden and visible sectors}.

\subsection{How thermal history of hidden and visible sectors enter in supercooled phase transitions \label{sec:3.3}}
In section \ref{sec:U1supercooled}, we have discussed some details of supercooled phase transition when assuming $\xi(T_v) =\zeta(T_h)= 1$. However, in most cases the temperature ratio varies significantly during and before the phase transition, especially for a decaying hidden sector, as discussed in the previous section. In this section, for a temperature dependent $\xi(T_v)$ or $\zeta(T_h)$, we will investigate how the analysis on supercooled phase transition could be modified. The percolation temperature, $T_{h,p}$, is defined as the point at which $71\%$ of the universe remains in the false vacuum, i.e.,
     \begin{align}
         P_f(T_{h,p}) = 0.71 \label{eq:Thp2}
     \end{align}
     where
     \begin{align}
         P_f(T_h)=\exp\left(-\frac{4\pi}{3}v_w^3\int_{T_h}^{T_{h,c}}dT'\frac{\Gamma(T_h')}{T_h'H^4(T_h')}\left(\int_{T_h}^{T_h'}dT_h''\frac{1}{H(T_h'')}\right)^3\right) \label{eq:Pff2}
     \end{align}
The Hubble parameter $H(T_h)$ is evaluated at the hidden sector temperature $T_h$:
\begin{align}
    H(T_h)= \sqrt{\frac{8\pi G}{3}\rho_{tot}} = \sqrt{\frac{8\pi G}{3}\left(\rho^v_{rad}(\zeta(T_h)T_h)+\rho^h_{rad}(T_h) +\rho^h_{vac}(T_h)  \right)} \label{eq:Hubble},
\end{align}

where $G$ is Newton’s gravitational constant,
\begin{align}
    \rho^v_{rad}(T_v) &= \frac{\pi}{30}g_{eff}^v(T_v)T_v^4,\\
    \rho^h_{rad}(T_h) &= \frac{\pi}{30}g_{eff}^h(T_h)T_h^4,\\
    \rho^h_{vac}(T_h) &\simeq \left.V_{\rm eff}^h(\phi,T_h) - T_h \frac{\partial V^h_{\rm eff}}{\partial T_h}\right|_{\phi = \phi_f}- \rho^h_{gs}, 
\end{align}
and  $\rho^h_{gs}$ in Eq.(\ref{eq:Hubble}) is the zero-temperature ground state energy density, and ${V_{\rm eff}^h(\phi,T_h)}$ is the effective potential for the hidden sector \cite{Athron:2022mmm}. 

\begin{figure}[h]
    \centering
    \includegraphics[width=0.33\linewidth]{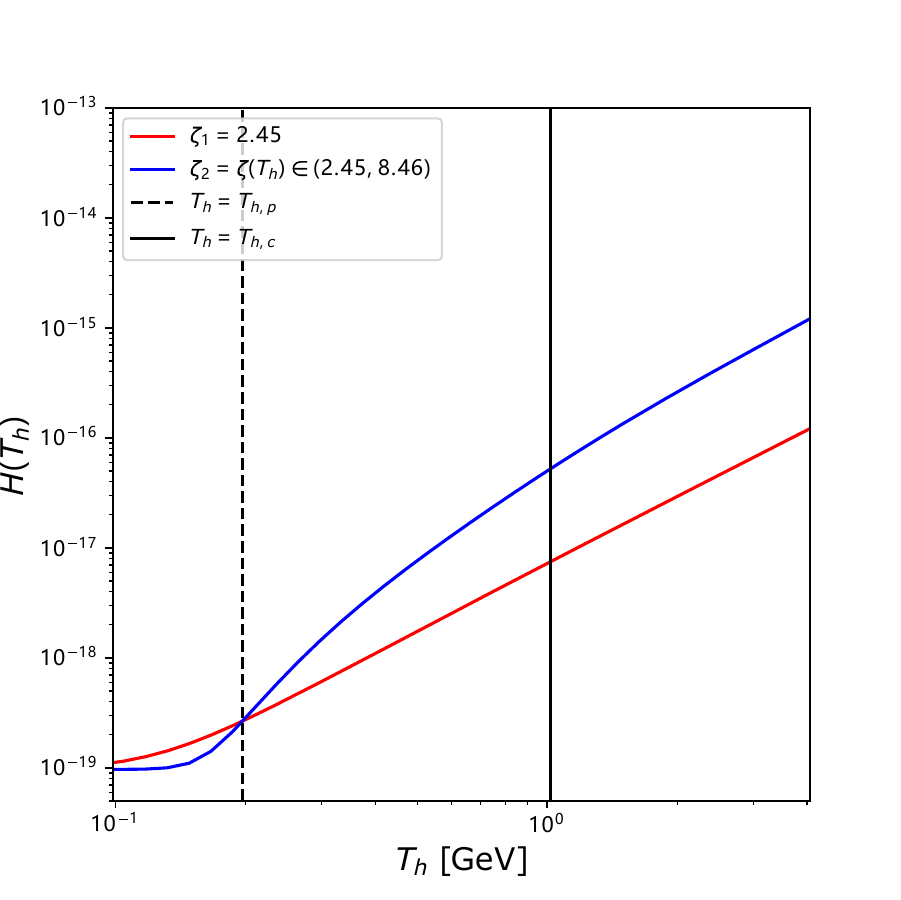}
    \includegraphics[width=0.31\linewidth]{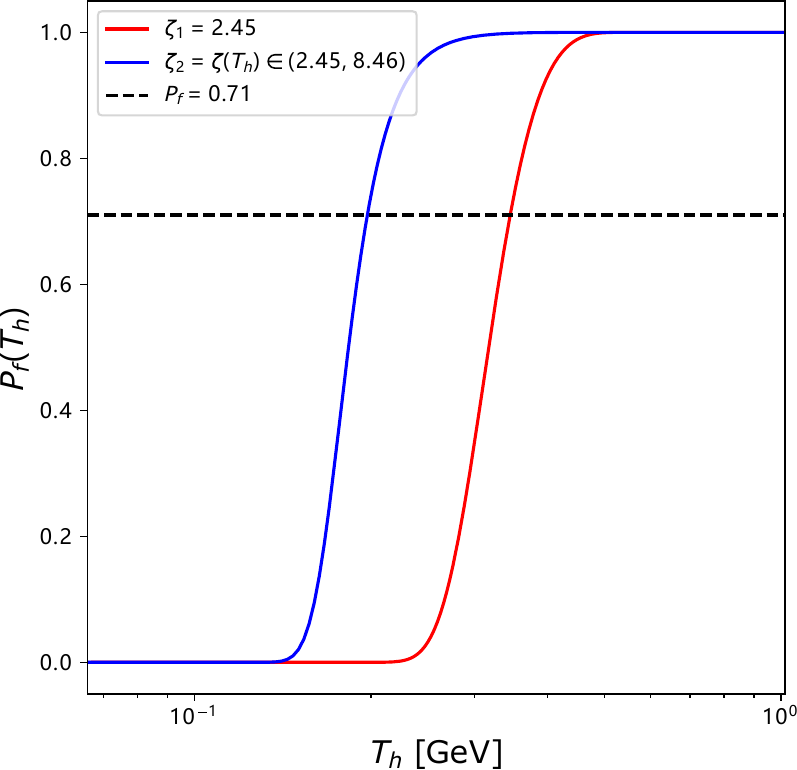}
    \includegraphics[width=0.32\linewidth]{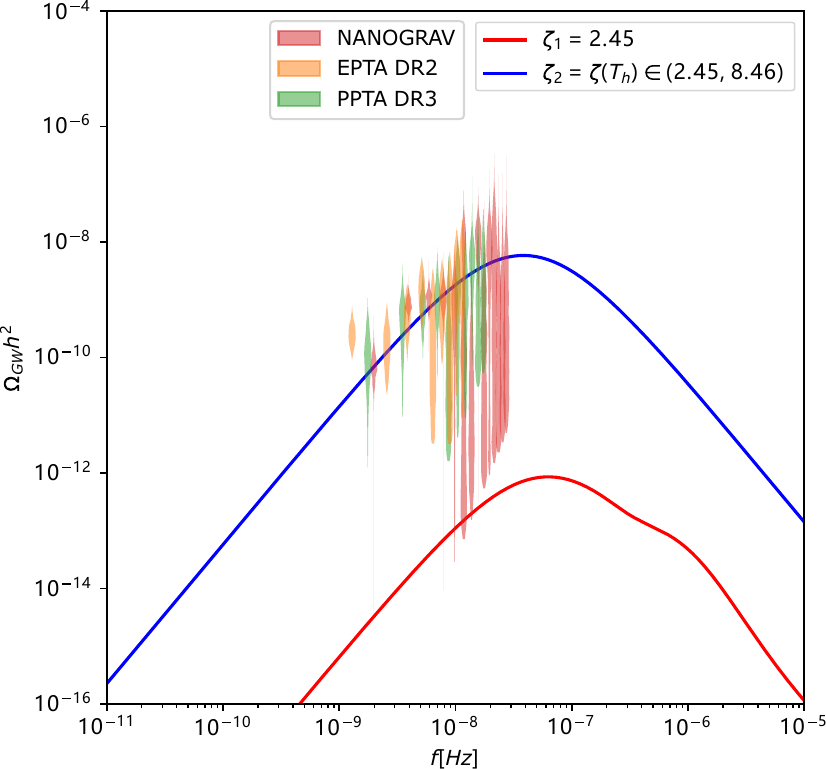}
    \caption{{ Illustration of the significant  effect of thermal histories of the visible 
    and the hidden sectors  on the Hubble parameter,
   percolation temperature and the gravitational power spectrum through their dependence on the parameter $\zeta(T_h)=T_v/T_h$.    
    Left Panel: The Hubble parameter
    $H(T_h)$ as a function of the hidden sector temperature $T_h$
    versus $T_h$ for two different choices of $\zeta$ where $\zeta$ is defined by 
    Eq.(\ref{eqzeta}).    
    The vertical dashed line marks the point  the percolation temperature for the 
    hidden sector $T_h=T_{h,p}$ and the solid line marks the critical temperature for the hidden sector $T_h = T_{h,c}$. Middle  Panel: $P_f(T_h)$ versus $T_h$ for two different choices of $\zeta$. The horizontal dashed line marks $P_f = 0.71$. Right Panel: Gravitational wave power spectrum $\Omega_{GW}h^2$ for two different choices of $\zeta$. Colored areas represent PTA signals. Benchmark model BP1 from Table. \ref{tab:benchmarks2} is applied for all three figures.}}
    \label{fig:H}
\end{figure}
The synchronous evolution {function}, $\zeta(T_h)$, appearing in Eq.~(\ref{eq:Hubble}), can modify the evolution history of the Hubble parameter and, consequently, affect the percolation temperature, since the Hubble parameter appears in the integral Eq.~(\ref{eq:Pff2}). For weakly supercooled or non-supercooled phase transitions, such effects are typically limited because $T_{h,p}/T_{h,c} \sim 1$, implying that the Hubble parameter does not change significantly during the small interval between $T_{h,p}$ and $T_{h,c}$. However, for strongly supercooled transitions, its influence becomes significant. To illustrate the Hubble parameter's sensitivity to temperature ratio evolution, we consider two different scenarios: $\zeta_1 = \text{const} = 1/\xi_p$ and $\zeta_2 = \zeta(T_h)$. In the first scenario, the temperature is constant, which is commonly suggested in most works in 
the literature.

 In the second scenario, we introduce an evolving temperature ratio. An important observation is that for these two scenarios, we have $\zeta_2 = \zeta(T_{h,p}) = 1/\xi_p = \zeta_1$. Thus, for weakly supercooled or non-supercooled phase transitions, these two scenarios are identical. 
We then plot $H(T_h)$ for these two scenarios on the left panel of Fig.~\ref{fig:H} using the benchmark model BP1. Here, we use dashed and solid lines to mark $T_h = T_{h,p}$ and $T_h = T_{h,c}$, which are the upper and lower limits appearing in the Integral Eq.~(\ref{eq:Pff2},\ref{eq:Rstar}). The figure shows that at $T_h = T_{h,p}$, the Hubble parameter is the same for both scenarios as discussed, while it can differ by a factor of 10 at $T_h = T_{h,c}$. In the middle panel, we have $P_f(T_h)$ versus $T_h$ for two different choices of $\zeta$. We observe a significant difference between these two cases, where the percolation temperatures differ by  100\%. This substantial change in $T_{h,p}$ significantly affects other parameters, including $\alpha_{tot},H_*$ and $R_*$. 
The effect of the temperature dependence is even more drastic on the \gw power spectrum.
The  right panel of Fig.~\ref{fig:H} shows the \gw power spectrum for the two different $\zeta$. We notice that they differ by a factor of  as much as $10^4$, which would seriously 
impact fits to the PTA signals.
\section{Computation of the gravitational wave power spectrum and fits to PTA data
\label{sec:GW}}
We carry out the analysis in two parts. In the first part we will present a set of benchmark
models and study the power spectrum in detail as a function of the frequency for the 
supercooled phase transition. Here we show that the \gw power spectrum allows a fit
to the PTA data in the range of frequencies observed by NANOGrav, EPTA and PPTA.
In the second part we will present a Monte Carlo analysis relevant for the PTA signal.
We discuss these below.

\subsection{Benchmark models for PTA signal}
We provide five benchmark points in Table \ref{tab:benchmarks2} that generate sufficiently strong GWs to match the PTA signal while satisfying other cosmological constraints. 
The relevant  parameters are listed in Table \ref{tab:benchmarkresults2}, and the resulting \gw power spectrum is shown in Fig.~\ref{fig:GWcurves}.
\begin{table}[h]
 \centering
 \small
     \begin{tabular}{llllllll|ll}
   \hline
    &  $m_q$& $g_x$&$f_x$ & {$\delta$($\times10^{12}$)}  & $\xi_0$ & $\mu_h$ & $\lambda_h$ & $m_{A}$ & $m_\phi$ \\ \hline
(BP1)&28.69&2.147&1.53e-03&6.948&0.118&0.906&0.2805&3.67&1.28\\
(BP2)&32.28&2.824&1.38e-05&11.387&0.185&8.023&0.7706&25.81&11.35\\
(BP3)&8.43&2.105&2.53e-04&8.573&0.116&0.262&0.2601&1.08&0.37\\
(BP4)&31.41&2.378&1.20e-03&8.472&0.133&0.337&0.4055&1.26&0.48\\
(BP5)&25.67&0.821&9.15e-03&5.543&0.231&8.195&0.0086&72.64&11.59\\
   \end{tabular}
         \caption{Benchmark points BP1-BP5 where the parameters  
   $m_q, g_x, f_x, \delta,\mu_h,\lambda_h$ are defined as in Eq.(\ref{bsm}) and Eq.(\ref{pot-hid})  and $\xi_0$ is the initial temperature ratio between two sectors. All masses are in units of GeV. 
   }
    \label{tab:benchmarks2}
\end{table}
\begin{table}[h]
\small
 \centering
     \begin{tabular}{llllll|lllll}
   \hline
    & $T_{h,p}$ & $T_{h,reh}$ & $T_{h,c}$ & $T_{v,p}$ & $\xi_p$ & $\alpha_{tot}$& $\alpha_{h}$& $\frac{\beta_*}{H_*}$& $\frac{(8\pi)^{1/3}v_w}{H_*R_*}$ & $\Omega_{DM}h^2$   \\ \hline
(BP1)&0.197&0.679&1.017&0.483&0.408&0.298&1769&-308.89&4.05&0.039\\
(BP2)&1.592&4.500&6.934&1.979&0.804&1.509&1096&-261.70&4.05&0.109\\
(BP3)&0.050&0.200&0.299&0.138&0.362&0.979&6397&-375.37&3.86&0.049\\
(BP4)&0.066&0.228&0.341&0.154&0.425&0.789&2992&-331.40&3.91&0.050\\
(BP5)&2.969&13.66&23.415&5.966&0.498&2.790&2497&-29.01&6.38&0.014\\
   \end{tabular}
    \caption{A display of the output of Table(\ref{tab:benchmarks2}) for the benchmarks BP1-BP5.
    $T_{h,p}$ and $T_{v,p}$ are the percolation temperatures for the hidden and the visible 
    sectors where $T_{h,reh}$ and $T_{h,c}$ are the reheat and the critical temperatures and
    $\xi_p$ is $\xi$ at the percolation temperature.  Further, $\alpha_{tot}$ and $\alpha_h$ are defined by
    Eq.(\ref{alphatot},\ref{Eq.alphastar}), $\beta_*/H_*$ is defined by Eq.(\ref{eq:betaoverH}), $R_*$ by
    Eq.(\ref{eq:Rstar})  and $\Omega_{DM}h^2$ is the dark matter relic density. All temperatures are in unit of GeV. {The benchmarks BP1-BP5 belong to Case 3.
   }   }
    \label{tab:benchmarkresults2}
\end{table}
\begin{figure}[H]
    \centering
    \includegraphics[width=0.45\linewidth]{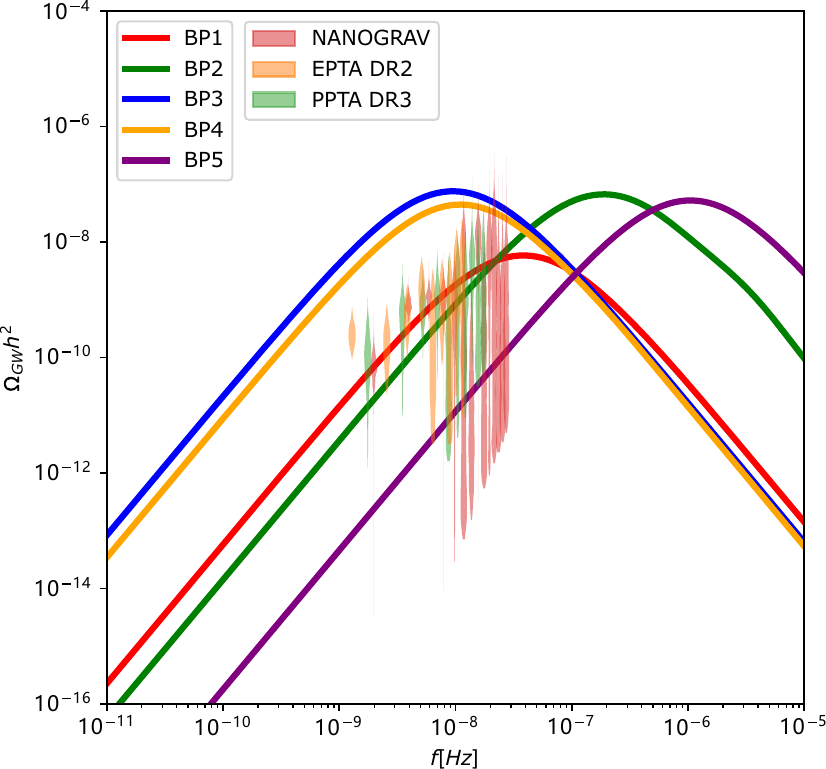}
    \includegraphics[width=0.45\linewidth]{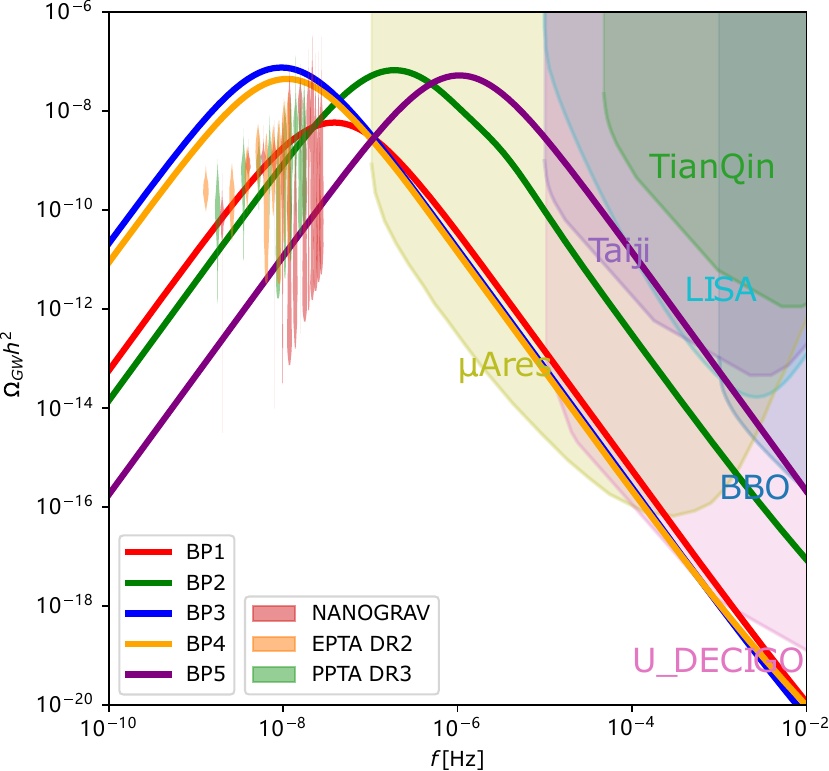}
    \caption{{Left Panel: The gravitational power spectrum $\Omega_{GW} h^2$ 
    for supercooled phase transitions as a function of the frequency $f$ in the range [$10^{-11}-10^{-5}$] [Hz] 
    for five benchmark points BP1-BP5. Colored areas represent PTA signals, given by NANOGrav~\cite{NANOGrav:2023gor}, PPTA~\cite{Reardon:2023gzh}, and EPTA~\cite{EPTA:2023fyk}. Right Panel: Detection regions from future space-based GW detectors are shown, including LISA \cite{LISA:2017pwj,Baker:2019nia,AmaroSeoane:2012km},  BBO\cite{Grojean:2006bp},  Decigo\cite{Kawamura:2006up}, Taiji\cite{Ruan:2018tsw}, TianQin\cite{TianQin:2015yph}, and $\mu$Ares\cite{Sesana:2019vho}. The right panel shows that supercooled phase transitions also can produce detectable signals in the higher frequency regions above NanoGrav.}}
    \label{fig:GWcurves}
\end{figure}
In Table \ref{tab:benchmarkresults2}, we list the signature temperatures, including the critical temperature $T_{h,c}$, reheating temperature $T_{h,reh}$, and percolation temperature $T_{h,p}$ in the hidden sector. We also provide the visible sector percolation temperature $T_{v,p}$, which is used when calculating the redshift. We observe that the temperature ratio during the phase transition, $\xi_p$, is generally large, which enables a strong phase transition, with $\alpha_{tot} \sim \mathcal{O}(1)$. For all benchmark points, the hidden sector becomes non-relativistic before BBN, thereby satisfying $\Delta N_{\rm eff}$ constraints. Our calculations of $\frac{\beta_*}{H_*}$ and $\frac{(8\pi)^{1/3}v_w}{H_*R_*}$ reveal that $\frac{\beta_*}{H_*}$ is negative for these benchmark points. While this would have excluded them under the previous constraint $\beta_*/H_* > 3$, these points are now valid since we employ the mean bubble separation $R_*$ instead. In the final column, we present the total relic density contributed by $\phi$ and $q$, calculated by solving the Boltzmann equations Eq.~(\ref{DE1},\ref{DE2},\ref{DE3},\ref{DE4}). These results demonstrate that our model can account for at least a portion of the dark matter content.
Fig. \ref{fig:GWcurves} displays the calculated \gw power spectra for all benchmark points. These spectra not only reach but in some cases exceed the observed PTA signal, providing strong potential for future observations. The right panel also illustrates the detection regions of future space-based \gw detectors, including LISA, BBO, Decigo, Taiji, TianQin, and $\mu$Ares. The power spectra from our benchmark points fall within these detection regions, suggesting that the model discussed here can be tested by these detectors in the future.\\

{ Here we wish to make a further remark regarding the experimental data and our analysis within 
supercooled phase transitions to generate a gravitational signal of size and frequency indicated by data.  
 We note that the experimental PTA observations are in the form of violins  which span an area with a significant spread in the $(f,\Omega_{GW} h^2)$ plane.  As observed by the referee the data has 
  gravitational signal mixed in with  errors (due to noise and systematics),  and it is hard to pin down the exact contours of the       signal within the area spanned by the violins. But the consistency of results from NanoGrav, PPTA and EPTA
       suggests 
      that the observations are not all purely due to errors and point to the possibility of a
      real gravitational signal.  Our approach is thus broad brush, i.e., we show that  supercooled phase
      transitions can generate gravitational signals which intersect a large part of the area populated by violins
      as illustrated by Fig.6 and Fig.8.  
      Thus the analysis presented in this work indicates that the model discussed in this work 
 can generate gravitational signals in the desired frequency range and strong enough to match the observational 
      signal when it is more accurately pinned down in the $(f,\Omega_{GW} h^2)$ space. Further, 
      the benchmarks BP1-BP5 and the 
 analysis of Fig. 6 show that a couple of models lie above the violins. This was done to show the robustness of our analysis. While many similar analyses predict outcomes that only reach the middle or lower parts of the violin, our analysis suggests the possibility that future data might reveal a gravitational signal even stronger than those observed so far. In such a case, the model we have discussed would have the potential to explain such a signal.}
   }\\

{
\begin{figure}[h]
    \centering
    \includegraphics[width=0.32\linewidth]{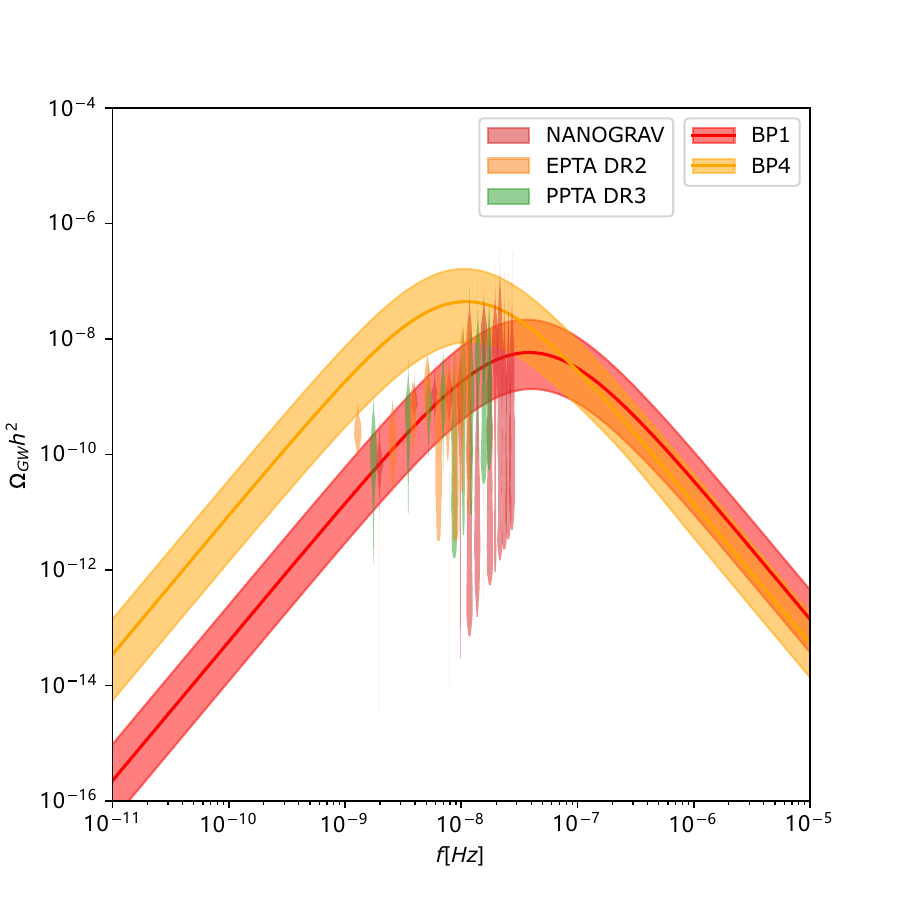}
    \includegraphics[width=0.32\linewidth]{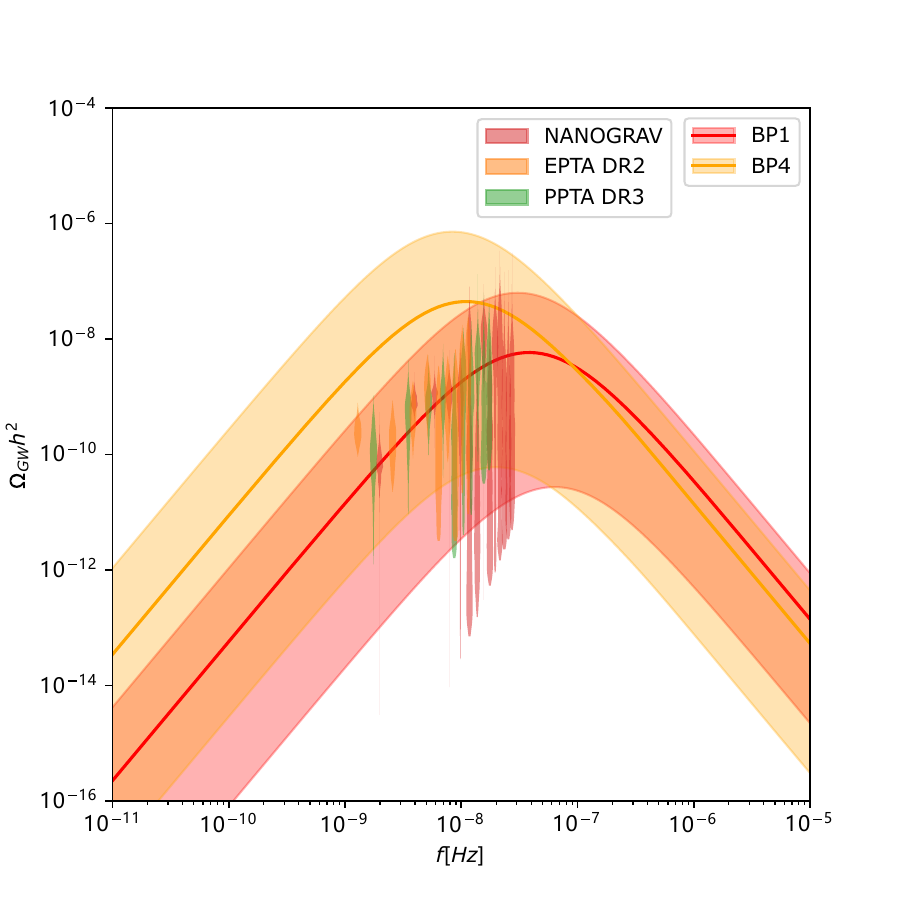}
    \includegraphics[width=0.32\linewidth]{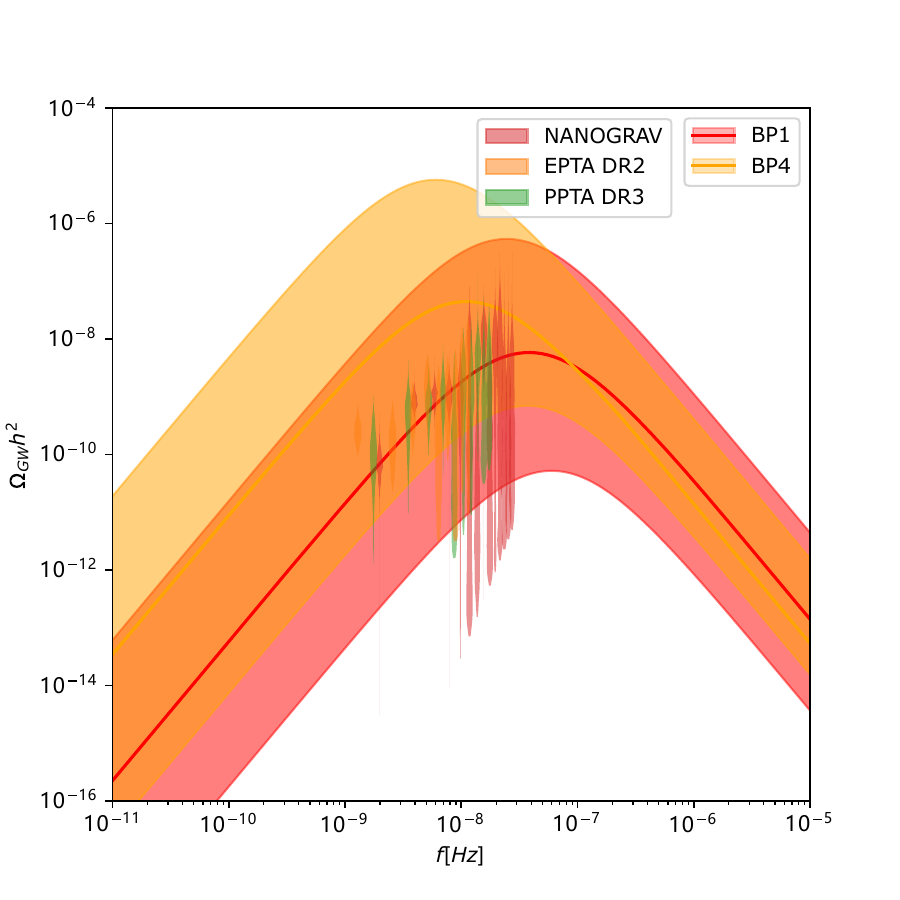}
    \caption{{
     An exhibition of sensitivity of the gravitational power spectrum as a function of frequency
     for variations of the inputs for benchmarks BP1 and BP4. 
   Left panel: An exhibition of the effect of  10\% variation in $m_q$, $f_x$, $\delta$, $\xi_0$, and $\mu_h$. Middle panel: An exhibition of the effect of  0.1\% variation in $g_x$ and $\lambda_h$. Right panel: An exhibition of the effect of variation in   $\xi^{\rm gauge}$        
       in  $R_\xi$ gauge fixing in the range $(-0.2,0.2)$. }}
    \label{fig:uncertainty}
\end{figure}
}

{
\subsection{Analysis of Theoretical Uncertainties}
This section quantifies the theoretical uncertainties associated with our numerical calculations and predictions. We employ a systematic parameter variation approach to evaluate the robustness of our results and identify the dominant sources of uncertainty. As established in Section~\ref{sec:U1supercooled}, the phase transition dynamics exhibit significant sensitivity to the coupling parameters $g_x$ and $\lambda_h$, which determine whether the phase transition is supercooled or not, whereas other model parameters, i.e.,  $(m_q, f_x, \delta, \xi_0, \mu_h)$,
 contribute only marginally to the overall uncertainty. Consequently, we present separate uncertainty analyses for these parameter groups.

For two benchmarks BP1 and BP2 
 shown in Fig. 6, we vary the weakly sensitive inputs $(m_q, f_x, \delta, \xi_0, \mu_h)$ by 
 $10\%$ and the critical couplings $(g_x, \lambda_h)$ by $0.1\%$, producing the uncertainty bands shown in Fig. 7. To assess the sensitivity of the effective potential to variations in $\xi^{\rm gauge}$
in the $R_\xi$ gauge-fixing we  vary $\xi^{\rm gauge}$ in the range $(-0.2,0.2)$ around the Landau gauge ($\xi^{\rm gauge}=0$). 
The analysis of the right panel shows that variations in $\xi$ can generate substantial changes
in the power spectrum. However, the variation in the power cannot be construed as 
a possible uncertainty of the Landau gauge prediction. The latter can only be assessed by its
comparison with an analysis using  gauge invariant effective potential
which includes Nielsen corrections.
Nonetheless the analysis of the right panel underscores the importance of Nielsen corrections 
that must be included for precision predictions. Note that our general conclusion that
supercooled phase transitions can generate low frequency gravitational waves which lie in 
the PTA observable range are still valid even accounting for uncertainties
as shown in Fig.~\ref{fig:uncertainty}.}

\subsection{Contribution to the GW Signal}
\begin{figure}[h]
    \centering
    \includegraphics[width=0.5\linewidth]{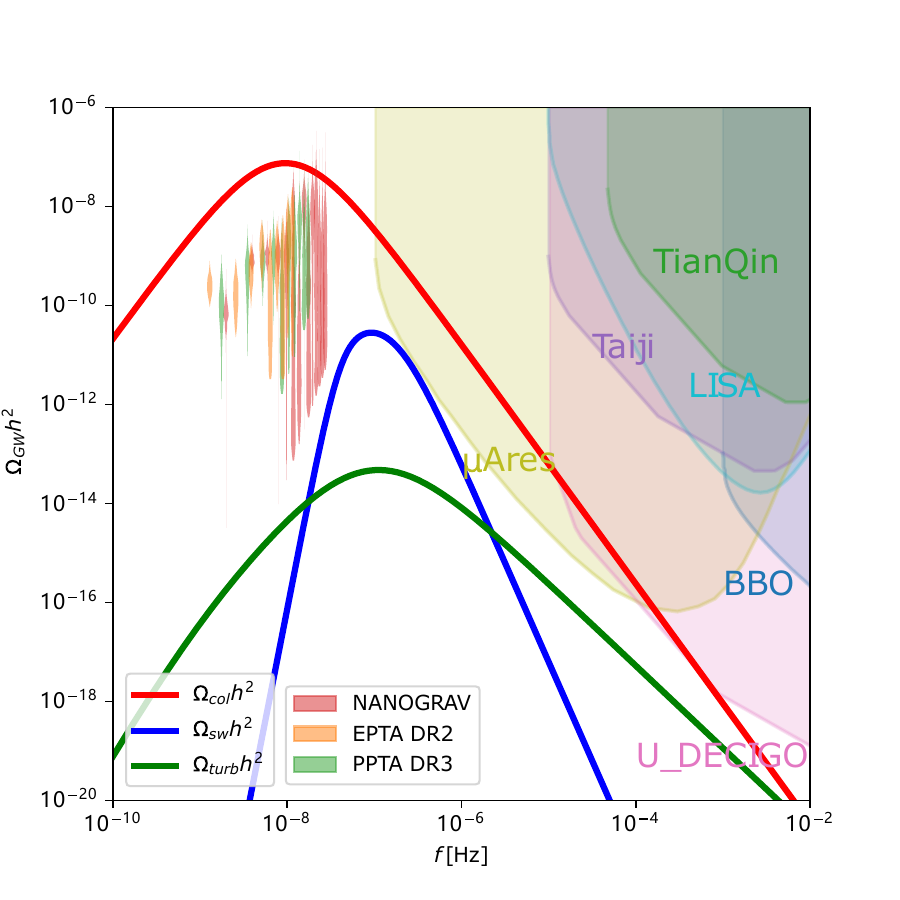}
    \caption{The three primary contributions to the GW signal for BP3: bubble collisions (red), sound waves in the plasma (blue), and turbulence in the plasma (green).
   The analysis indicates that the dominant contribution to the gravitational waves 
    comes from bubble collisions while the contributions from sound waves and turbulence
    are essentially negligible in the nano-Hz region.}
    \label{fig:GWcontribution}
\end{figure}


{
There are three primary contributions to the GW signal: bubble collisions, sound waves in the plasma, and turbulence in the plasma. If a true vacuum bubble generated during the phase transition continues to accelerate to a velocity close to the speed of light (\(v_w \simeq 1\)), this is referred to as the \textrm{runaway scenario}. In this case, most of the energy is stored in the bubble walls, and GWs are predominantly generated by bubble collisions. Conversely, if the bubble reaches a terminal velocity \(v_w < 1\), it is termed the \textrm{non-runaway scenario}. Here, most of the energy resides in the plasma, and GWs are primarily produced by sound waves and turbulence. The distinction between the runaway and non-runaway scenarios depends on whether the friction exerted by the plasma is comparable to the pressure difference driving the bubble expansion \cite{Caprini:2015zlo}. A detailed analysis of this is provided in Appendix \ref{sec:7.1}.
For all BPs, we generally find that \(\alpha_{h,\infty} < \alpha_h\), indicating that they all correspond to runaway scenarios. As shown in Fig.~\ref{fig:GWcontribution}, the dominant contribution to the GW signal comes from bubble collisions.
}


\subsection{Monte Carlo analysis of model points covering the PTA signal}
\begin{figure}[h]
    \centering
    \includegraphics[width=0.4\linewidth]{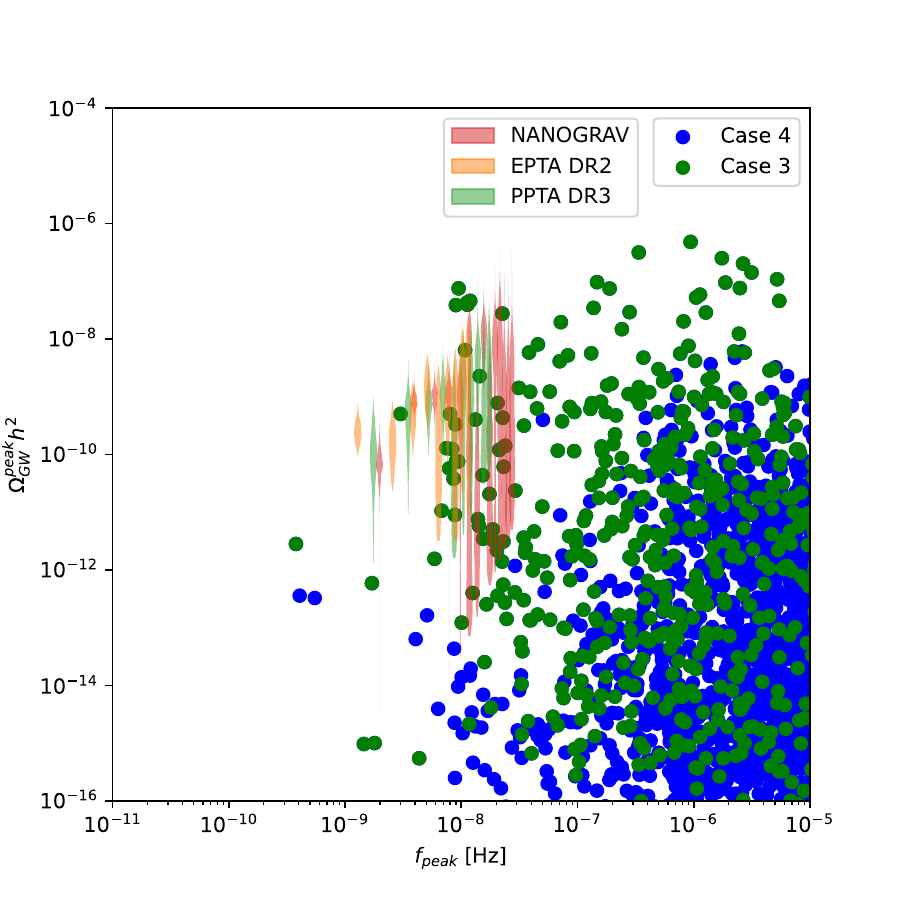}
    \includegraphics[width=0.43\linewidth]{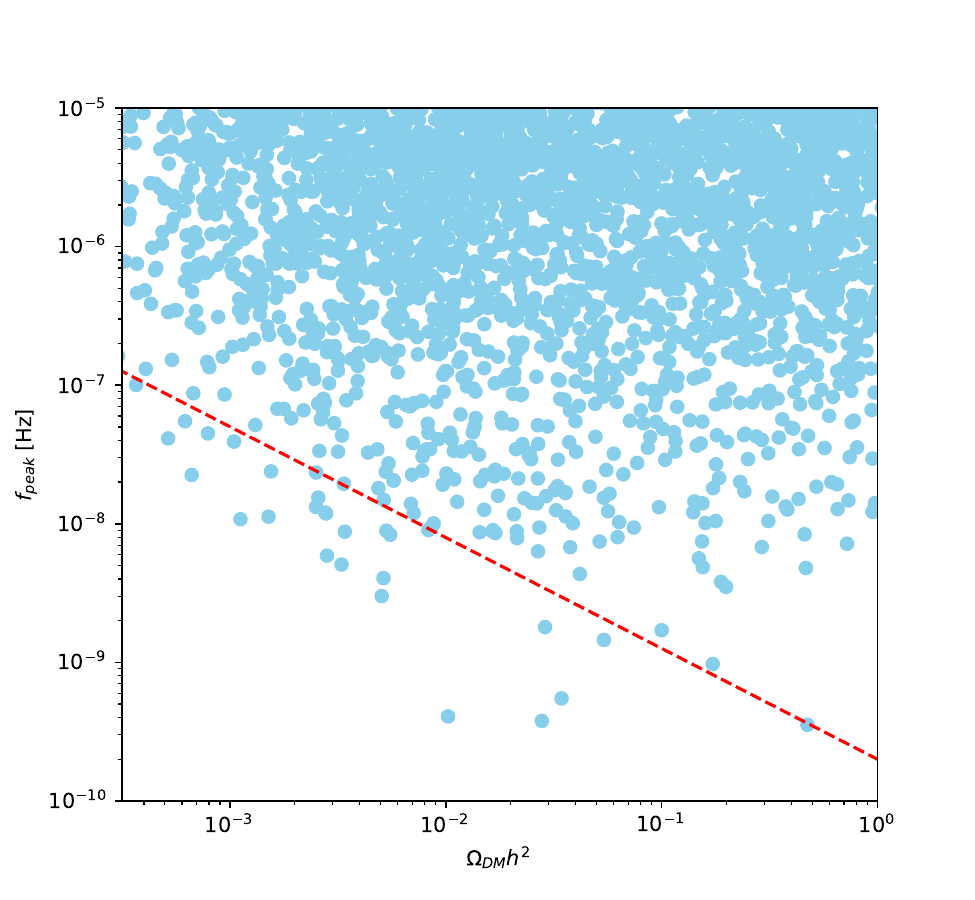}         
    \caption{Left panel:
    {Result of Monte Carlo analysis showing gravitational power spectrum peaks  
    for {two different supercooled phase transition scenarios: Case 3 and Case 4.}
     Green points represent Case 3 and blue points are for Case 4. All points shown satisfy the cosmological constraint on $\Delta N_{\text{eff}}$ and dark matter relic density requirement ($0.001 < \Omega_{\text{DM}}h^2 < 0.12$). The sampling was performed along the boundary of the ultracooled region with small fluctuations. 
     {Right panel: 
     Result of Monte Carlo analysis showing peak frequency vs the dark matter relic density in the 
      frequency regions of NanoGrav, PPTA and EPTA data. The scatter plot exhibits a trend where 
      larger dark matter density favors a smaller  peak frequency. }  }}
    \label{fig:MC}
\end{figure}

Fig.~\ref{fig:MC} presents the result of a Monte Carlo analysis performed on 7 free parameters: $m_q$,$g_x$,$f_x$, $\delta$, $\xi_0$, $\mu_h$, $\lambda_h$. 
To conserve computational resources, we conducted the Monte Carlo sampling along the boundary of the ``ultracooled" region with small fluctuations. Each point in the figure represents the peak of a \gw power spectrum curve. All displayed points correspond to successfully completed phase transitions that satisfy both the cosmological constraint on $\Delta N_{eff}$ and the dark matter relic density requirement ($0.001 < \Omega_{DM}h^2 < 0.12$). 
In the left panel of Fig.\ref{fig:MC} the points are color-coded, with green and blue points corresponding to Case 3 and Case 4, respectively, as discussed in section \ref{sec:U1supercooled}.
The results demonstrate that numerous Case 3 {models} readily achieve PTA signal levels. In contrast, virtually none of the Case 4 {models} reach PTA signal levels while simultaneously satisfying all the cosmological constraints.
  {In right panel of Fig. \ref{fig:MC} we give a scatter plot of peak frequency vs $\Omega_{DM} h^2$. Here one finds
   a trend in that the large dark matter relic density tends to lead to smaller gravitational wave frequency.
    Since both the relic density and the gravitational wave frequency depend on multiple factors, a 
    analytical explanation of this trend is not so obvious and needs further investigation.}

\section{Gravitational Wave Energy Density in PTA signal region\label{sec:ED}}
In this section we discuss the energy transfer to gravitational waves during the phase transition. An approximation of the energy density ratio is given by \cite{Athron:2023xlk}:
\begin{align}
    \frac{\rho_{GW}}{\rho_{tot}} \sim \kappa^2\left(\frac{\alpha}{1+\alpha}\right)\left(\frac{\beta}{H}\right)^{-2}.
\end{align}
However, as discussed in Section \ref{sec:U1supercooled}, $\beta$ may not be an appropriate parameter to describe the transition rate of a supercooled phase transition. Here we compute the energy density ratio numerically. The gravitational wave
 power spectrum before redshift is defined as:
\begin{align}
    \Omega_{GW} = \frac{1}{\rho_{tot}}\frac{\partial \rho_{GW}}{\partial\ln{f}}
\end{align}
With $\Omega_{GW}$ given in Section \ref{Appendix}, we can determine the ratio:
\begin{align}
    \frac{\rho_{GW}}{\rho_{tot}} = \frac{\int d\rho_{GW}}{\rho_{tot}} = \int_{-\infty}^\infty \frac{\Omega_{GW}}{f}df.
\end{align}
Furthermore, we have $\alpha_{tot}\simeq \frac{\rho_{vac}}{\rho_{tot}}$. Using these relations, we can calculate $\frac{\rho_{GW}}{\rho_{vac}}$.
\begin{table}[h]
    \centering
    \begin{tabular}{c|ccc}
         No.& $\frac{\rho_{GW}}{\rho_{tot}}$(units:$10^{-3}$) & $\alpha_{tot}\simeq\frac{\rho_{vac}}{\rho_{tot}}$ & $\frac{\rho_{GW}}{\rho_{vac}}$(units:$10^{-3}$)\\
         \hline
         BP1& 0.31 & 0.30 & 1.10\\
         BP2& 3.30 & 1.51 & 2.18\\
         BP3& 2.83 & 0.98 & 2.89\\
         BP4& 1.73 & 0.79 & 2.19\\
         BP5& 2.39 & 2.79 & 0.85
    \end{tabular}
    \caption{Energy density ratios for different benchmark points (BP1-BP5) of Table    (\ref{tab:benchmarks2}).  The table gives the ratio of the gravitational wave energy density to the total energy density ($\frac{\rho_{GW}}{\rho_{tot}}$), the approximate ratio of the vacuum energy density to the total energy density ($\alpha_{tot}$), and the resulting ratio of the gravitational wave energy density to the vacuum energy density ($\frac{\rho_{GW}}{\rho_{vac}}$).}
    \label{tab:energyratio}
\end{table}

Table \ref{tab:energyratio} presents the ratios for five benchmark points given in Table \ref{tab:benchmarks2}. As shown, for {all the benchmark points shown}, only about {(0.1-0.3)}\% of the vacuum energy is transferred to the gravitational wave energy. Thus, the energy conservation assumption expressed in Eq.(\ref{Eq.reheat}) remains valid.

\section{Conclusion \label{sec:conclusion}}
In this paper, we explored the possibility that the nano-Hertz \gw signal observed by the recent PTA experiments arise from a supercooled first-order phase transition  within a hidden sector featuring a spontaneously broken \(U(1)_X\) gauge symmetry.
We emphasize in our analysis the use of mean bubble separation \(R_{*}\), rather than the inverse timescale \(\beta_{*}\), in characterizing supercooled phase transitions. In Fig.~\ref{fig:Randbeta}, we demonstrated how naively adopting \(\beta_{*}/H_{*}\) for strongly supercooled scenarios can lead to substantial errors. Here we show that a large portion of the parameter space that would otherwise appear excluded by the commonly used condition \(\beta_{*}/H_{*} > 3\) is in fact viable once \(R_{*}\) is used in place of \(\beta\).

In addition, the main new feature of our analysis over existing works is that we have taken into account thermal histories of coupled hidden and visible sectors evolving their temperature ratio  \(\xi(T_{v})\) or alternately \(\zeta(T_{h})\) in a synchronous fashion.
 We highlight in Fig.~\ref{fig:H} that a nontrivial evolution of \(\zeta(T_{h})\) can significantly affect the Hubble parameter during the transition {(see the left panel of Fig.(\ref{fig:H}))}, thereby modifying the percolation temperature \(T_{h,p}\) and the mean bubble separation \(R_{*}\). In particular, the longer integration range between \(T_{h,c}\) and \(T_{h,p}\) in a supercooled scenario amplifies the impact of how \(\xi\) and \(\zeta\) evolve. 
 {The sensitivity to $\xi$ is also reflected in the percolation temperature as exhibited
 in the middle panel of Fig.(\ref{fig:H}).
 Further, ignoring the $\xi$  evolution can lead to discrepancies of up to four orders of magnitude in the predicted \gw power spectrum as exhibited  the right panel of Fig.(\ref{fig:H}).} 
The results of our analysis are illustrated through several benchmark points, each of which satisfies important cosmological and astrophysical constraints and gives an observable 
\gw power spectrum. We presented these spectra in Fig.~\ref{fig:GWcurves} and demonstrated that hidden sector FOPTs can generate \gws in the PTA frequency range with sufficient power to match current observations. Further, some benchmark models predict signals within the reach of future \gw experiments such as LISA, Taiji, and others.

In summary, our study shows that {with a more complete analysis which properly takes into account the subtleties discussed in this work}, the supercooled hidden sector phase transitions can naturally generate \gw backgrounds that explain the current PTA data. We have provided a systematic treatment of the supercool
phase transitions and demonstrated that both the mean bubble separation \(R_{*}\) and the synchronous temperature ratio \(\xi(T_{v})\) and \(\zeta(T_{h})\) play crucial roles in the 
\gw power spectrum analysis. Consequently, the analysis presented here offers a robust framework for further investigation of the cosmic first order  phase transitions and for 
further exploration of their role in the analysis of stochastic  \gw backgrounds in 
$[10^{-9}-10^{-2}]$Hz frequency range with levels of power spectrum accessible to 
 current and future \gw detectors.\\  
 
 \noindent
{\bf Acknowledgments}\\
The research was supported in part by the NSF Grant PHY-2209903.

\section{Appendix: Details of GW Power Spectrum \label{Appendix}}
 A number of factors enter in the \gw power spectrum and have been discussed in 
 depth in the literature. We give here a brief summary of some of the elements
that enter prominently in the analysis. Thus \gw power spectrum involves the hydrodynamics of the bubble nucleation includes the  efficiency factor $\kappa$  and the bubble wall velocity 
 $v_w$. Further, the \gw power spectrum receives contributions from bubble collision,
 from sound waves and from turbulence. We discuss these further below.
 
\noindent
\subsection{ Hydrodynamics, Efficiency Factor and
Bubble Wall Velocity \label{sec:7.1}}
{In the main text of the paper}, we have discussed many aspects of the supercooled phase transition in a hidden $U(1)_X$ model, including temperatures $T_{h,p}$ and $T_{h,reh}$, transition strengths $\alpha_{tot}$ and $\alpha_h$, and transition rates $\frac{\beta_*}{H_*}$ and $\frac{1}{H_*R_*}$. {In the analysis of \gws}  generated by a first-order phase transition (FOPT), there are two remaining parameters to consider, which {enter in the} hydrodynamics of bubble expansion: the efficiency factor $\kappa$ and bubble wall velocity $v_w$ \cite{Bodeker:2017cim,Espinosa:2010hh,Giese:2020rtr,Giese:2020znk,Ai:2021kak,Ai:2024uyw}. Here, the transition coefficient $\kappa(c_{s,f}^2,c_{s,t}^2,\alpha,v_w)$ is calculated using $\alpha_h$ rather than $\alpha_{tot}$. Detailed calculations of $\kappa$ can be found in \cite{Giese:2020znk,Giese:2020rtr}, and we utilize their provided Python code for numerical calculations. The kinetic energy fraction is then given by
\begin{align}
    K = \frac{\Delta\Bar{\theta}(T_{h,p})}{\rho_{tot}}\kappa(c_{s,f}^2,c_{s,t}^2,\alpha_h,v_w) \simeq \frac{\alpha_{tot}}{1+\alpha_{tot}}\kappa(c_{s,f}^2,c_{s,t}^2,\alpha_h,v_w)
\end{align}

To determine the bubble wall velocity, we must first classify whether the bubble expansion is runaway or non-runaway \cite{Caprini:2015zlo}. This classification is made using a critical phase transition strength $\alpha_{h,\infty}$, defined as:
\begin{align}
    \alpha_{h,\infty} =  \frac{(T_{h,p})^2}{\rho^h_{\rm rad}(T_{h,p})}\left(\sum_{i=\text{bosons}}n_i\frac{\Delta m_i^2}{24} + \sum_{i=\text{fermions}}n_i\frac{\Delta m_i^2}{48}\right)
\end{align}

In the runaway regime, where $\alpha_{h,\infty}<\alpha_h$, we have:
\begin{align}
    \kappa_\phi = 1-\frac{\alpha_\infty}{\alpha_h},\quad \kappa_{\rm sw} = \frac{\alpha_\infty}{\alpha_h}\kappa(\alpha_\infty,c^2_{s,f},c^2_{s,t},v_w),\quad v_w = 1
\end{align}

Conversely, in the non-runaway regime, where $\alpha_{h,\infty}>\alpha_h$, we have:
\begin{align}
    \kappa_\phi = 0,\quad \kappa_{\rm sw} = \kappa(\alpha_h,c^2_{s,f},c^2_{s,t},v_w) 
\end{align}
For the non-runaway case, $v_w$ requires further determination. Following the method described in \cite{Ai:2023see,Ai:2024uyw}, we calculate $v_w(\alpha_h,c_{s,f}^2,c_{s,t}^2,\Phi_p)$, where:
\begin{align}
    \Phi_p = \frac{w_t(\phi_t,T_{h,p})}{w_f(\phi_f,T_{h,p})}, \quad w^h_i(\phi_i,T_h) = p^h_i(\phi_i,T_h)  + \rho^h_i(\phi_i,T_h) 
\end{align}
with $p^h_i$ and $\rho^h_i$ given in Eqs.~(\ref{eq:phi}, \ref{eq:rhohi}). Our analysis shows that for any supercooled phase transition events capable of {producing a PTA signal, the 
$\alpha_h$ is generally too large to yield deflagration or hybrid solutions. Thus all PTA events 
end up as  detonation type} with 
\begin{align}
v_w \approx v_J = c_{s,f}\left(\frac{1 + \sqrt{3\alpha_h(1+c_{s,f}^2(3\alpha_h-1))}}{1+3\alpha_hc_{s,f}^2}\right),
\end{align}
{where}  $v_J$ is the Chapman-Jouguet velocity. Thus, taking $v_w = v_J$ serves as a reasonable approximation, at least within the context of the hidden $U(1)_X$ model under investigation.

\subsection{Details of the gravitational power spectrum}
Gravitational waves produced during the early universe undergo redshifting. To correctly match the gravitational waves we observe today with those produced during the early universe, we need to consider this effect. For a hidden sector phase transition, the power spectrum $\Omega^0_{GW}(f_0)$ at the current temperature $T_{v,0}$ from the power spectrum $\Omega_{GW}$ obtained at the temperature $T_{v,reh}$ is given by
\begin{align*}
   \Omega^0_{GW}(f_0) &= \mathcal{R}_\Omega \Omega_{GW}\left(\frac{1}{\mathcal{R}_f} f_0\right),~~
    \frac{1}{\mathcal{R}_f}  = \frac{a_0}{a} = \left(\frac{h_{eff}(T_{v,reh})}{h^{0}_{eff}}\right)^{1/3}\left(\frac{T_{v,reh}}{T_v^0}\right)\\     
    \mathcal{R}_\Omega &\equiv \left(\frac{a}{a_0}\right)^4\left(\frac{H}{H_0}\right)^2 \simeq 2.473\times 10^{-5}h^{-2} \left(\frac{h^{0}_{eff}}{h_{eff}(T_{v,reh})}\right)^{4/3}\left(\frac{g_{eff}(T_{v,reh})}{2}\right)\\
    g_{eff}(T_{v,reh}) &= g_{eff}^v(T_{v,reh}) + g_{eff}^h(T_{v,reh})\xi(T_{v,reh})^4\\    
    h_{eff}(T_{v,reh}) &= h_{eff}^v(T_{v,reh}) + h_{eff}^h(T_{v,reh})\xi(T_{v,reh})^3\\
    h^{0}_{eff} &\simeq 3.91 
\end{align*}
where $T_v^0 = 2.35\times10^{-13}$GeV is today's temperature. Since the reheating occurs within the hidden sector, the visible sector temperature remains unaffected. Consequently, the final GW power spectrum is not significantly impacted even in the presence of strong reheating, unlike the case discussed in \cite{Athron:2023mer}.
The computations of the gravitational wave power spectrum have been carried out 
by a number of authors ~\cite{Hindmarsh:2016lnk,Hindmarsh:2019phv,Gowling:2022pzb,Hindmarsh:2017gnf,Cai:2023guc,Ghosh:2023aum,Caprini:2010xv,Caprini:2009yp,Caprini:2015zlo,Weir:2017wfa}.  
These results are collected in Appendix E of \cite{Athron:2023mer} which we utlize in the
analysis of this work. The \gw power spectrum includes contributions from bubble collision, from sound waves, and from turbulence. Their relative contributions are discussed in detail
in Fig. (\ref{fig:GWcontribution}) where one finds that the PTA signal is dominated by the
\gw arising from bubble collision.



\end{document}